\documentclass[fleqn,usenatbib]{mnras}
\usepackage{float}
\usepackage{amsmath}
\usepackage{amssymb}
\usepackage{rotating}
\usepackage{tabularx}
\usepackage{gensymb}
\usepackage{longtable}
\usepackage{lscape}
\usepackage{xr}

\include{journaldefs}
\usepackage[font=footnotesize,labelfont=bf]{caption}

\begin{document}

\title[\bf OB-star proper motions in the far Carina Arm]{\bf 
Proper motions of OB stars in the far Carina Arm
}
\author[J. E. Drew et al]{
{\parbox{\textwidth}{J. E. Drew$^{1}$\thanks{E-mail: j.drew@ucl.ac.uk}, M. Mongui\'o$^2$, N. J. Wright$^3$}
}\\ \\
$^{1}$Department of Physics \& Astronomy, University College London, Gower Street, London, WC1E 6BT, UK \\
$^{2}$Institut de Ci\`encies del Cosmos (ICCUB), Universitat de Barcelona (IEEC-UB), Mart\'i Franqu\`es 1, E08028 Barcelona, Spain\\
$^3$Astrophysics Group, Keele University, Keele, ST5 5BG, UK\\
}

\maketitle

\begin{abstract}
In large scale maps of the Galactic disc, the Carina Arm stands out as a clear spiral feature, hosting prominent star clusters and associations rich in massive stars.  We study the proper motions of 4199 O and early B most likely in the far Carina Arm, at distances mainly in excess of 4~kpc from the Sun, within the sky region, $282^{\circ} < \ell < 294^{\circ}$ and $-3^{\circ} < b < +1^{\circ}$ (Galactic coordinates). The sample is constructed by extending an existing blue-selected catalogue, and cross-matching with Gaia EDR3 astrometry.  The observed pattern of proper motions is modulated into a saw-tooth pattern, with full amplitude approaching 1 mas yr$^{-1}$, recurring roughly every 2--3 degrees of longitude (200--300 pc at the median OB-star distance of 5.8 kpc). Kinematic perturbation of underlying circular rotation is most likely present.  The data also reveal a moving group containing $>50$ OB stars at $\ell \sim 286^{\circ}$, $b \sim -1^{\circ}.4$ behind the main run of the far arm.  An analysis of relative proper motions is performed that yields an incidence of runaway O stars of at least 10\% (potentially $>20$\% when full space motions become available).  To map where runaways have run away from, we set up simulations for the region that assume linear trajectories and test for trajectory impact parameter in order to identify likely ejection hot spots. We find the method currently gives good results for times of flight of up to $\sim$4~Myr.  It shows convincingly that only NGC 3603 and Westerlund 2 have ejected OB stars in significant numbers.  Indeed, both clusters have experienced intense spells of ejection between 0.6--0.9 and 0.5--0.8 Myr ago, respectively.  
\end{abstract}



\section{Introduction}

The internal environments of spiral galaxies, including our own, are critically shaped by their OB-star populations.  The collective action of their winds, their output of ionising radiation, and the production of supernova explosions are all important factors in determining galactic evolution.  Our growth in understanding how these massive-star populations work has been hindered by the fact of their infrequency in the solar neighbourhood.  In the wider Milky Way, particular challenges have come from their kiloparsec-scale distances, their obscuration by interstellar dust, and from the bias towards finding them within bright clusters. As a result, the balance of roles between clustered and field OB stars presently remains uncertain.

An aspect of this is the continuing interest in how much of the 'field' population of massive stars can be accounted for by runaways.  There is a growing appreciation that the ejection of massive stars from clusters has an impact on galactic evolution and galaxy-scale outflow, as shown recently by \cite{Andersson20}. Runaway OB stars are thought to be produced through one of two mechanisms \citep{Blaauw1961}. The first is the binary supernova scenario in which the lower-mass companion to a star undergoing core collapse is ejected. The second is dynamical ejection, in which a close binary interacts with a single star or another binary system, leading to the ejection of one or more stars \citep[e.g.,][]{Fujii2011,Perts2012,Oh2016}. The latter mechanism is expected to produce the fastest runaways \citep[e.g.,][]{Renzo2019}. The dominant cluster property driving dynamical ejections is its density \citep[as well as the primordial binary fraction,][]{Oh2016}, and thus the ejection times of runaway stars may be used to trace the historical properties of the cluster.

Given the obstacles in the way of finding and analysing the Milky Way's OB stars, much of the recent progress has come from the VLT Flames Tarantula Survey of OB stars in the lower metallicity environment of the Magellanic Clouds \citep[][and subsequent papers]{Evans2011}, where extinction is low and large populations are accessible within a limited sky area. In a different series of papers, Oey and collaborators have used studies of the Small Magellanic Cloud to build a case for a high proportion of field stars being ejections from clusters \citep[see, most recently,][]{Oey2018, DJones2020}.

With the advent of digital wide field surveys conducted on ground-based telescopes, now enhanced by the availability of space-based astrometry from the Gaia mission, it is possible to begin to exploit the maximally resolved Galactic populations of OB stars on scales extending well beyond that of individual clusters \citep[cf. recent work by][]{Wright2020,Zari2021}. 
This work is also timed to capitalise on the release of the early third data release of Gaia astrometry \citep[EDR3,][]{GaiaEDR3}, in a case study of the far Carina Arm, where there are thousands of massive O and early B stars available over tens of square degrees of sky.  Our emphasis will be on what can be learned from the much improved proper motions, both collectively and individually.  

This study follows on from three pieces of work.  The first of them by \cite{MMS2017} provides a large photometrically-identified and spectroscopically-tested catalogue of Carina OB stars that we re-use and add to here.  Since this appeared, we have already made use of it in searching the environs of the massive young clusters, NGC 3603 and Westerlund 2, for runaway O stars \citet[areas of 1.5$\times$1.5 deg$^2$ in both cases]{Drew2019, Drew2018}.  Proper motions from Gaia DR2 astrometry were key to both searches.   Here we take a much wider view that spans a sky area of $4\times12$ square degrees and ranges over a distance range running from 4 to $\sim$8 kpc -- a volume increase of $\sim$20 relative to each of the previous studies.  This provides a vision that zooms out to capture the massive star demographics of an entire spiral arm segment roughly 500 parsecs across.  
Our focus is on the far arm.  The near arm (containing the Carina Nebula and associated clusters) is omitted for the reason that coverage of the brighter stars associated with it would be significantly incomplete, whilst also contributing relatively few stars.


The contents of this paper are organised as follows.  The extraction of a far Carina Arm sample of 4199 OB stars from a cross-match between an extended version of the \cite{MMS2017} catalogue and Gaia EDR3 is described in Section~\ref{sec:sample}.  Sections~\ref{sec:structure1} and \ref{sec:structure2} provide an overview of the OB-star sample and of the Carina far arm proper motions. An outcome from this is the uncovering of a group of over 50 OB stars that appears to be behind the main run of the Carina arm at Galactic longitude $\ell \sim286^{\circ}$ -- this is presented in Section~\ref{sec:other}.  We then move onto consideration of the vector point diagram of the sample proper motions, and two ways of examining runaway-candidate statistics in the region (Section~\ref{sec:PMstructure}).  The next step, in Section~\ref{sec:runaways},
is to develop and characterise the search method used by \cite{Drew2019} to find runaways from the massive young cluster, NGC 3603, and apply it across the entire region, in order to expose any further sites of OB-star ejection within it  and to update the DR2-based results of \cite{Drew2018} and \cite{Drew2019} to EDR3. 
The paper ends with conclusions and a brief discussion (Section~\ref{sec:discussion}).

\section{Construction of the sample}
\label{sec:sample}

The basis for the sample is the Carina OB-star catalogue of \citet[herafter MS17]{MMS2017}.  The first step in preparing this catalogue was to select from VPHAS$+$ photometry obtained prior to 2015.  The selection was made using the $(u-g)$ versus $(g-r)$ diagram, where a cut was applied that should separate off stars with spectral types hotter than B3. To refine this, the blue photometric selection was then cross-matched with 2MASS (to add $JHK$ to the optical $ugri$ bands) and run through a Monte Carlo Markoff chain analysis that identified the combination of main-sequence model-atmosphere magnitudes and extinction best reproducing the observed spectral energy distribution.  Crucial to the success of this procedure was the inclusion of the $u$ band.  Parameters from the fitting included estimates of effective temperature, visual extinction ($A_0$), extinction law index ($R_V$), and the posterior-distribution median $\chi^2$ as a measure of fit quality.  In line with there being three degrees of freedom, MS17 separated 'good' OB star candidates from the more questionable by requiring $\chi^2 < 7.82$. The spectroscopic test of the catalogue reported by MS17 indicated low levels of contamination by unwanted later-type stars: 97\% of 276 objects were confirmed as O and early B stars.   

The MS17 catalogue misses some of the region running between Galactic longitudes 282$^{\circ}$ and 294$^{\circ}$, and between Galactic latitudes $-3^{\circ}$ and $+1^{\circ}$.   To complete the full 48 sq.deg sky area, we have applied the same selection to those parts of the region that were not ready for analysis by MS17.


Starting from the extended MS17 catalogue, a bound of $\chi^2 < 30$ is applied in order to eliminate the clear interlopers in the original $(u-g,g-r)$ photometric selection.  A tail of objects in the range $7.82 < \chi^2 < 30$ is retained because experience has shown they are frequently classical Be or Wolf-Rayet stars (see MS17).  The next cut applied is to remove all stars with $\log{T_{\rm eff}} < 4.3$, to achieve a sample focused on O and the earlier B stars.  With the cooler stars eliminated, 6493 stars of the original MS17 list and 898 objects from the added sky area remain under consideration.

\begin{figure}
\begin{center}
\includegraphics[width=0.8\columnwidth]{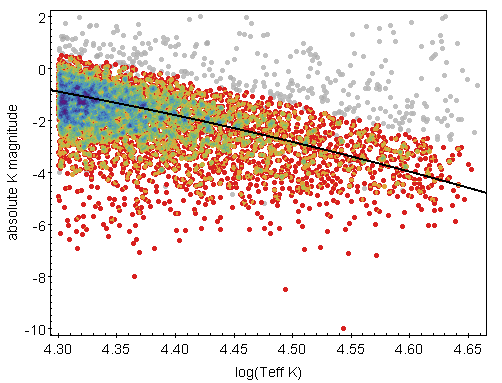}
\caption{
Absolute K magnitude as a function of $\log{T_{{\rm eff}}}$ for the 4812 stars making up the sample after the Gaia EDR3 crossmatch and the rejection of $\log{T_{{\rm eff}}} < 4.3$, $\chi^2 > 30$ and $K > 14$ stars.  The stars in grey do not survive into the final sample of 4199 stars (see text).  The coloured stars are the survivors, where a blue colour implies maximum density in the plot and red, the lowest.  The black line shows the predicted dependence of $M_K$ on $\log{T_{{\rm eff}}}$ on the upper main sequence. 
}
\label{fig:MK_v_logteff}
\end{center}
\end{figure} 

\begin{figure}
\begin{center}
\includegraphics[width=0.8\columnwidth]{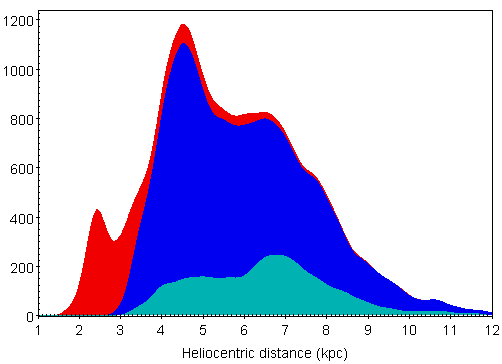}
\caption{
Distribution of heliocentric distances $D_{2.0}$ computed via the EDSD method using a length scale of 2 kpc and a systematic parallax offset of 0.03 mas.  The final sample of 4199 stars is in blue, while the red distribution (underneath) comprises the 4812 stars remaining after the Gaia EDR3 crossmatch and cut on $K$ magnitude.  The final sample (both O and B stars) can be seen to be largely free of near-arm stars producing the peak between 2 and 3 kpc.  The O star distribution (with $\log{T_{{\rm eff}}} > 4.45$) is shown in cyan. Kernel density estimation (KDE) has been applied here, using a gaussian kernel with $\sigma = 0.2$~kpc. 
}
\label{fig:distance_hist}
\end{center}
\end{figure} 

\begin{figure*}
    \centering
    \includegraphics[width=2.1\columnwidth]{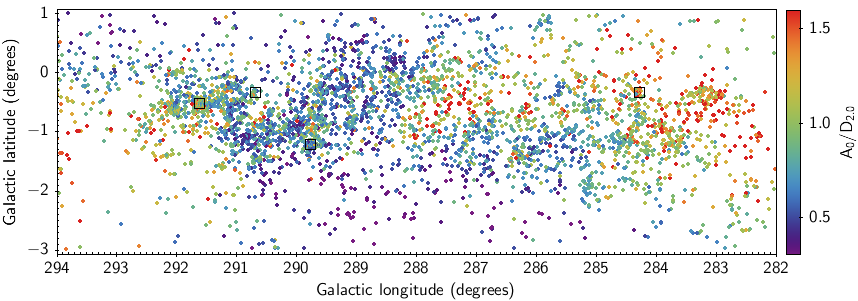}
    \caption{Map of the 4199 stars making up the final far-arm sample.  The candidate OB stars are coloured according to their measured visual extinction per kpc of distance (as inferred from the EDR3 parallax).  Working from left to right in the plot, black boxes identify the positions of NGC 3603, the WRA 751 cluster, the association around the O5 supergiant LSS 2063, and Westerlund2.
    }
    \label{fig:map}
\end{figure*}

The sample was then cross-matched with the Gaia EDR3 database, setting a maximum accepted position offset of 0.5 arsecond and requiring the reduced unit weight error to be less than 1.4.  This produced a list of 6195 stars in which the median cross-match angular separation was 0.124 arcsec. Next, a condition that the 2MASS $K$ magnitude should not exceed 14 was also imposed.  This limit is aimed at removing sub-luminous lower mass stars (sub-dwarfs mainly), along with stars that might be far beyond the Carina Arm. Note that a B2-3 main-sequence star, towards the bottom end of the range of interest has a visual absolute magnitude between $-1$ and $-2$ \citep{Zorec1991}.  Hence at 10 kpc and a visual extinction of 8 magnitudes (the high end for our list) the expected K magnitude would be $K \sim 14$, assuming $V-K \simeq -0.7$ \citep[see table in][]{Kenyon1995}.  Objects fainter than this are unlikely to be the sought-after far-arm massive stars. This cut reduced the list to 4812 stars. 

As the aim is to build a statistical description of the far Carina Arm, stars more likely to be in the near arm need to be weeded out.  This demands a distance estimate: we compute 
these via the EDSD method presented by \cite{Luri2018}. To do this we need to adopt an appropriate parallax offset, and a length scale that plausibly mimics the drop off in stellar density with distance.  The parallax offset we use (0.03 mas) is motivated by a result of \cite{Lindegren21b} who have examined the EDR3 astrometry of Galactic-plane clump giants at longitudes, adjacent to Carina, ranging from $\ell = 265^{\circ}$ to $275^{\circ}$: they find these red objects still require this larger offset (compared with the network of bluer quasars outside the plane, now indicating $-$0.019 mas).  On grounds of both proximity on the sky and the similarly red colours, it makes sense to use the same offset as for the clump giants.  The choice of length scale has been guided by the distribution of absolute $K$ magnitude, $M_K$, as a function of effective temperature (as estimated from photometry).  We estimate $M_K$ for every sample star via the expression:
\begin{equation}
M_K = K - 5\log(D_L) - 5 - 0.11A_0    
\end{equation}
where $K$ is the 2MASS $K$ magnitude, $D_L$ the inferred distance for length scale $L$, and $A_0$ is the catalogued visual extinction. We use the $K$ band here to bear down on the absolute magnitude error contributed by the extinction correction.  A fixed scaling of $A_K$ to $A_0$ is applied for simplicity: the range of variation in $R_V$ among the OB stars only affects the scale factor at the level of $\sim0.01$).   We find that $L = 2.0$ kpc returns a trend with effective temperature that is consistent with expectation for mainly main-sequence O and early B stars \citep[see Fig.~1, the predicted line drawn in black is based on Padova data,][]{Bressan2012}.  The dependence on $L$ is modest in that a switch to the slightly inferior length scales of either 1.5 or 2.5 kpc shrinks and expands the distance scale by around 10 percent.

On the basis of the computed $D_{2.0}$ values (see Fig.~\ref{fig:distance_hist}), it is evident that the near Carina Arm dominates for $D_{2.0} \lesssim 3$~kpc.   So we reject those stars returning most probable distances less than 3.0 kpc, along with a smaller number of stars with good parallaxes (errors 10 percent or less) at distances up to 3.3 kpc.  This removes 397 stars, leaving 4415 stars. 
To shore the sample up further, we eliminated stars with $M_K$ more than 1.5 magnitudes fainter than the minimum main sequence prediction for the estimated effective temperature.  The contribution of $A_K$ to the $M_K$ error budget is $\sim$0.05 mag, while distance uncertainty contributes from $\sim$0.2 to over 0.5 mag.  Stars are thus lost to the sample either because of larger random parallax errors or because they are faint later-type interlopers.  In choosing 1.5 mag as the maximum allowable offset fainter than trend, we are cutting at 1.1$\sigma$ of the scatter.  After these stars are removed, the final far-arm sample is reduced to 4199 stars.

A map of the positions of the 4199 stars retained is shown in Fig.~\ref{fig:map}.  
A list of them all is supplied in machine readable form in supplementary materials.  For each object, photometric magnitudes are given along with estimates of key stellar parameters deduced from VPHAS$+$ and 2MASS photometry, along with distances and Galactic-coordinate proper motions obtained from Gaia EDR3 astrometry.  Details on the parameters provided are given in Appendix~\ref{full_list}.  For the objects appearing in the MS17 catalogue, their names are retained and have the form VPHAS-OB1-nnnnn, where nnnnn is a serial number ($<$ 15000, ordered by Galactic longitude).  When individual objects are referred to individually in this study, the shortened form of name, \#1$-$nnnnn, will be used.  The newly added objects have names of the form VPHAS-OB2-nnnnn, where the lowest serial number is 15000, and the ordering is again by Galactic longitude.  Where names of these objects are needed in text, they will be shortened to \#2$-$nnnnn.

Constructed as described above, the final sample splits into 930 candidate O stars, and 3269 B stars.

The fraction of stars with $7.82 < \chi^2 < 30$ is 12.0\% (501 objects).  Strikingly, but as expected, a high proportion of them (47\%, or 235 stars) are flagged as likely emission line stars.  Among the better-fit $\chi^2 < 7.82$ group, the emission line candidates are much less common, only numbering 144 out of 3698 stars (just 4\%).  This is why it is helpful to extend the permitted $\chi^2$ range: without raising the bound to $\chi^2 = 30$, the representation of Be and Oe stars in the sample would be well below the Milky Way norm of $\sim$10\% (see MS17).  In this sample, 379 (9\%) qualify as emission line stars, based on the criterion that the ($r - H\alpha$) excess is 0.1 above the mean OB-star reddening line in the ($r - H\alpha$, $r - i$) diagram.  This roughly corresponds to a lower limit on H$\alpha$ emission equivalent width of 10~\AA .

\section{Overview of the far-arm sample}

\begin{figure}
    \centering
    \includegraphics[width=0.8\columnwidth]{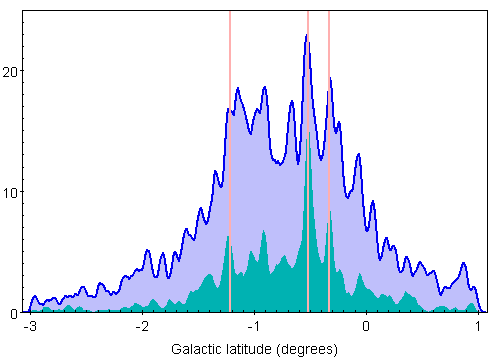}
    \caption{The Galactic latitude distributions of the B candidates (blue, and behind) and O candidates (cyan, in front). Both are shown in KDE form with an Epanechnikov smoothing.  The vertical lines are drawn at the latitudes of Westerlund 2 (least negative), NGC 3603 and the LSS 2063 association (most negative).
    }
    \label{fig:latitude}
\end{figure}

\subsection{The two-dimensional distribution and distances}
\label{sec:structure1}

There is broad acceptance that NGC 3603, the outstanding massive young cluster in Carina, is located in the far Carina Arm at a distance of 7 to 8 kpc \citep[][give 7.2, 7.7, 6.9 and 7.6 kpc]{Melnick1989, Nurnberger2002, Sung2004, Melena2008}. The other well-studied massive young cluster in the area is Westerlund 2.  Its distance was once controversial, but the range cited has latterly narrowed to between 4 and 6 kpc \citep[][respectively obtain 4.2, 4.7 and 6.0 kpc]{Vargas2013, MMS2015, Dame2007}.  The two shorter distances rest on stellar photometry, while the highest is a kinetic distance from molecular data.  These two clusters are marked in the sky map of the 4199 stars provided as Fig.~\ref{fig:map}, respectively at $\ell = 291^{\circ}.62$ and $\ell = 284^{\circ}.27$

Other clusterings are apparent. Prominent among them is the LSS~2063 group of OB stars, first recognised in the optical by MS17.  It is roughly centered on ($289^{\circ}.77$, $-1^{\circ}.22$), has a mass in the region of 8000~M$_{\odot}$ and is better viewed as an association since it is about 20 pc across \citep{MMS2017}, if -- as seems likely -- it is about as far away as NGC 3603. A further cluster in this part of the far arm surrounds the position of the luminous blue variable, WRA~751 \citep{vanGenderen01}.  These are also marked in Fig.~\ref{fig:map} and will be considered briefly in section 5 along with NGC 3603 and Westerlund 2.

The OB stars in the region prefer negative Galactic latitudes, with 90\% of the distribution falling inside the latitude range $-2^{\circ}.0 < b < 0^{\circ}.5$.  Indeed the histogram of Galactic latitudes, presented in Fig.~\ref{fig:latitude}, shows this is a property shared by the O stars and early B stars. Again, NGC 3603, Westerlund 2 and the LSS 2063 association are evident as peaks superposed on the broader distribution. 

We now turn to how the sample is distributed in heliocentric distance according to Gaia EDR3 parallaxes.  The distribution of inferred distances $D_{2.0}$ was shown in Fig.~\ref{fig:distance_hist}.  The expectation would be that the B stars should favour a nearer distance range than the O stars, since the underpinning photometric selection is magnitude-limited.  This is seen in that the O stars are relatively more frequent beyond $\sim$6 kpc. Despite this, the O stars present slightly better parallaxes than the B stars on account of being brighter -- for them, the median ratio of parallax to parallax error is 5.94, while it is 4.77 for B stars.  The median distance for the O stars is 6.55 kpc, as compared with 5.60 kpc for the B stars. For the full sample the median is 5.84~kpc. Increasing (decreasing) the adopted length scale in the EDSD parallax inversion will stretch (compress) the distance scale, with only modest impact on the O to B distance offset. 

An impression of the pattern of random errors can be obtained from the top panel of Fig.~\ref{fig:pmsvlong}, where stars in the main latitude band, $-2^{\circ}.0 < b < +0^{\circ}.5$, are plotted.  At the longitudes of Westerlund 2 and NGC 3603 there are prominent 'fingers of god', along which likely cluster stars are strongly scattered in distance. The growth in random error on individual objects is from around 1 kpc at $\ell \sim 284^{\circ}$ near Westerlund 2, to over 2 kpc at $\ell \sim 291^{\circ}$ near NGC 3603. 

For the clusters the distance errors can be beaten down. For Westerlund 2, the mean parallax among 26 stars, positioned within 1 arcmin of cluster centre is 0.1946$\pm$0.0073 mas.  Direct inversion of the parallax (whilst still applying a global offset of 0.03 mas) yields a distance of 4.45 ($+0.15,-0.14$).  Analogously, for NGC 3603, from 34 stars within 1 arcmin of the centre, the mean parallax is 0.1068$\pm$0.0068 mas, inverting to give 7.31 ($+0.38, -0.35$) kpc. The much reduced error in each case, and correspondingly much larger ratio of the parallax to the error,  means that inversion via the EDSD method, requiring the adoption of a finite scale length, gives very nearly the same, if marginally lower, distances.  The estimates obtained in this way for both clusters sit in the midst of the distances based on stellar photometry cited at the start of this section.

The data in Fig.~\ref{fig:pmsvlong} hint at two main groupings of the OB population, with a rift of lower stellar density between them at longitudes running from $\sim 287^{\circ}$ to $\sim289^{\circ}$, particularly at $5 < D_{2.0} < 6$ kpc: below it and at lower longitudes, the inferred distances tail off by $D_{2.0} = 6$ kpc, while at higher longitudes most stars are found in the range $5 < D_{2.0} < 9$~kpc.  The origin of this behaviour is presently unclear.  In part, it may be an effect of raised extinction among stars located behind the Carina Nebula in the near arm at around $\ell \sim 287^{\circ}.5, b \sim -0^{\circ}.5$ (increased extinction is visible here in Fig.~\ref{fig:map}). Extra extinction has the consequence that the OB-star faint detection limit is brighter in this locale, leading to lowered catalogued stellar density. A more comprehensive explanation would be that there is a rupture in the run of the far arm that results a bimodal distance distribution in this longitude range.  A clearer picture may emerge when better parallaxes become available.

\subsection{Insights from proper motions on the structure of the far Carina Arm}
\label{sec:structure2}

\begin{figure}
\begin{center}
\includegraphics[width=1.0\columnwidth]{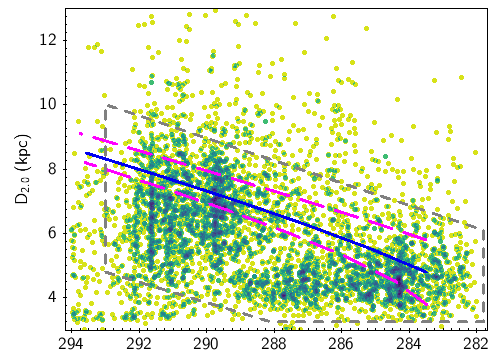}
\includegraphics[width=1.0\columnwidth]{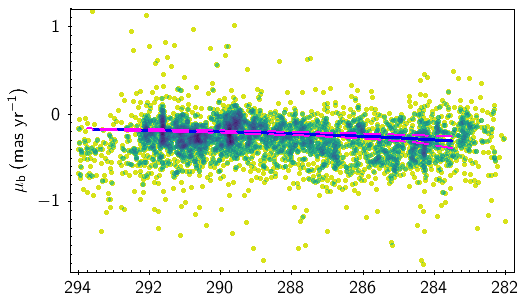}
\includegraphics[width=1.0\columnwidth]{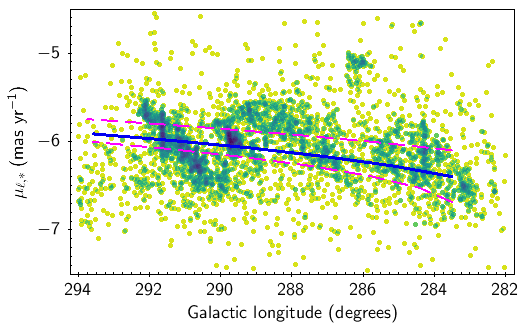}
\caption{Astrometric properties of the selected OB stars in the latitude range $-2^{\circ}.0 < b < 0^{\circ}.5$, as a function of Galactic longitude (3811 stars). The distances, $D_{2.0}$, in the top panel have been computed from EDR3 parallaxes by the EDSD method as described in text. The superposed grey-dashed box is used in section 4.2 to delimit a more conservative far-arm sample. The middle and lower panels show the proper motion (PM) components re-expressed in Galactic coordinates.  The parameter exhibiting the most pronounced structure is the longitude component of the PM, $\mu_{\ell,*}$ in the bottom panel.  Elongated strips of stars ('fingers of God') created by e.g. NGC 3603 at $\ell \sim 291^{\circ}.6$ are evident in all three panels.  The superposed blue line in each case is the trend predicted for a smooth spiral, computed for a pitch angle of 10$^{\circ}$, with a reference point at $\ell = 283^{\circ}.5$ at a heliocentric distance of 4.8 kpc.  The magenta dashed lines are the predictions that apply when the reference point is moved closer and further away by 1 kpc.  
}   
\label{fig:pmsvlong}
\end{center}
\end{figure}

Fig.~\ref{fig:pmsvlong} also shows how the Galactic longitude and latitude components of proper motions, $\mu_{\ell,*}$ and $\mu_b$, depend on Galactic longitude, for the stars within the densely populated $-2^{\circ} < b < +0.5^{\circ}$ belt.  The better precision of the proper motion data means they are more informative than the parallax-based distances.

Recently-born massive OB stars are expected to show little-to-no component of motion perpendicular to the Galactic Plane, as they are expected to conform more than (older) lower mass stars to purely circular motion.  Hence any latitudinal proper motion should simply reflect, on average, the Sun's motion out of the plane of the disk.  This component of the Sun's space motion is believed to be $+7\pm0.5$ km~s$^{-1}$ \citep[][and references cited]{BHandG2016}.  It would induce an apparent proper motion, $\mu_b$, of $-0.252\pm0.018$ mas yr$^{-1}$ in an object whose motion lies in the mid-plane of the disk at a heliocentric distance of 5.84 kpc, the median for the full sample. The results in the middle panel of Fig.~\ref{fig:pmsvlong} are consistent with this to within the errors on the Sun's motion: the median $\mu_b$ is $-0.269$ mas yr$^{-1}$.  It is also in accord with expectation that the bulk of the sample does not obtrude across the zero line (the infinite distance limit for no motion out of the Galactic plane). 

To aid our understanding, we can compare the longitude distribution of both $\mu_b$ and $\mu_{\ell,*}$ with the trend expected for an idealised spiral arm.  This is computed assuming a pitch angle of 10$^{\circ}$, applying the finding of \cite{Reid2019} that this angle gives a reasonable match to the outer Carina Arm as mapped in CO.  For present purpose, we adopt a reference longitude of $\ell = 283^{\circ}.5$ within the model arm, at a heliocentric distance of 4.8 kpc.  This places it within the far arm, clear of the CO tangent longitude of $\ell \simeq 282^{\circ}$ \citep[see e.g.][]{Dame2007} and in among the first density peaks in our sampling.  In all three panels of Fig.~\ref{fig:pmsvlong} the trend given by the illustrative model arm is shown in blue.  These trend lines are not extended inwards to capture the turn at the tangent and the near arm, both because they would quickly exit the domain occupied by the sample stars and because the full specification of the Carina Arm remains unsettled (cf Reid et al 2019, Dame 2007 and Vall\'ee 2014).  

First, we return to $\mu_b$. 
The trend line in the middle panel of Fig.~\ref{fig:pmsvlong}  
is accompanied by dashed magenta lines that delimit the range in outcome on varying the distance to the reference point by $\pm$1 ~kpc.  The data and the model arm are broadly consistent, even if the data exhibit significant dispersion (the standard deviation on the mean for the plotted sample is 0.295 mas yr$^{-1}$, or $\sim 8$ km s$^{-1}$ at the median sample distance).  The plot vividly demonstrates that the dependence of $\mu_b$ on distance, compared to the dispersion, is slight.  

The situation is quite different regarding the longitude distribution of $\mu_{\ell,\ast}$.
We note that the median random error in $\mu_{\ell,\ast}$ is 0.026 mas yr$^{-1}$, while the systematic error is now of order 0.010 mas yr$^{-1}$ \citep{Lindegren21a}.  This yields a total error that is small compared to the variations seen in the bottom panel of Fig~\ref{fig:pmsvlong}, giving a standard deviation across the region of 0.58 mas yr$^{-1}$ (or $\sim 16$ km s$^{-1}$ at median distance).  Maintaining for the moment the assumption that OB stars, on average, follow circular orbits within the Galactic disk, $\mu_{\ell,*}$ would then change monotonically with increasing distance from around $-6.7$ mas yr$^{-1}$ at $D \simeq 4$~kpc and $\ell \sim 283^{\circ}$, to $-5.7$ mas yr$^{-1}$ at $D \simeq 8$~kpc, and $\ell \sim 292^{\circ}$.
The blue trend line superimposed on the $\mu_{\ell,*}$ data 
in Fig.~\ref{fig:pmsvlong} uses the disk rotation law presented by \cite{Eilers19}.  This law shows a slight decline in rotation speed with increasing Galactocentric radius ($-1.7$ km s$^{-1}$ per kpc, with a circular speed of 229 km s$^{-1}$ on the Solar Circle).  Had a flat or slightly rising law been deployed instead, the outcome would change only subtly.  Indeed, the outcome only begins to shift noticeably if the adopted circular speed changes by $\sim$5~km s$^{-1}$, while the overall trend remains essentially the same.  In a highly averaged sense, the data for the OB stars follow the expected trend -- but the variation around the ideal case is substantial and exhibits interesting structure.


The $\mu_{\ell,*}$ distribution appears to break up into three roughly-defined ranges, $283^{\circ}.0-285^{\circ}.5$, $286^{\circ}.5-289^{\circ}.0$, and $290^{\circ}-292^{\circ}$.  Within each segment there is an upward slope in $\mu_{\ell,*}$ with increasing longitude.  This is particularly marked in the last of them, where the dispersion about the upward trend is less.  Taken together, the pattern across the longitude range is one of a skewed wave, seen to approach full amplitude 1 mas yr$^{-1}$ in the $>290^{\circ}$ segment. At the distances of 6--8~kpc prevailing at these longitudes, this difference translates to an in-sky velocity contrast of 33 $\pm$ 6 km s$^{-1}$, if interpreted as a purely kinematic perturbation.  Alternatively the pattern seen may arise from modulated line-of-sight distance, in the case that the OB stars follow unperturbed circular motion in the mean.

With sufficiently precise parallax distances the second option can be verified easily. Presently it is evident that the organised $\mu_{\ell,*}$ variation is not mirrored at all in the distances inferred from the EDR3 parallax data.  But there remains plenty of room to doubt individual parallaxes are yet good enough on scales of 4--8 kpc to decide the matter.  We can deduce that the character of the $\mu_{\ell,*}$ modulation present -- in the case the OB stars conform with disk rotation -- demands ordered line-of-sight distance excursions within each longitude segment of 2--3 kpc, whilst the $\Delta\ell$ range within each segment is equivalent to 0.2--0.3 kpc in the plane of the sky.  Essentially, the plot of distances inferred from $\mu_{\ell,*}$ against longitude exhibits the same structure as the bottom panel of Fig.~\ref{fig:pmsvlong}. The emergent picture is then of a highly corrugated far arm.  

A check on this interpretation is available using the cluster, Westerlund 2, for which we formed the mean parallax and inferred distance in section~\ref{sec:structure1}.  The relatively precise result obtained can be compared with the distance inferred from the mean longitudinal proper motion for the same stars ($-5.921\pm0.027$ mas yr$^{-1}$).  This mean maps onto a distance of 6.0 ($+$0.2,$-$0.1) kpc, in close agreement with the kinematic distance obtained by \cite{Dame2007}.  But it is appreciably longer than the parallax result, 4.45 ($+$0.15,$-$0.14) kpc.  However there are important systematics to examine before concluding on this.  The proper motion distance has a dependence on the adopted Galactic rotation speed: if this is varied by $\pm 10$ km s$^{-1}$ relative to the \cite{Eilers19} value, allowing it to range from $\sim$240 down to $\sim$220 km s$^{-1}$, the distance then obtained ranges from roughly 5.5 to 6.5 kpc.   The parallax-based estimate is based on 26 stars falling within a 1-arcmin radius circle: according to equation 26 in \cite{Lindegren21a} the EDR3 angular covariance at such small separations creates an uncertainty of up to $\pm 0.0265$ mas.  Resetting the bounds on the measured parallax accordingly causes the inferred distance to range from 3.98 up to 5.05 kpc. This allows the gap between the two estimates to narrow to $\sim$0.5 kpc, if the systematics are so large and combine appropriately.  Complete elimination of the discrepancy appears improbable.


Given the above issues, it is unlikely that the proper motion data can be explained under the constraint that the OB stars rigorously follow disk rotation. But it remains open that some kinematic perturbation combines with some distance modulation.  Support for this can be taken from the positive $\mu_b$ gradients just discernible in the longitude ranges, $286^{\circ}.5-289^{\circ}.0$, and $290^{\circ}-292^{\circ}$.  However, the gradients are small and uncertain, relative to the amplitude of scatter.

Beyond $\ell \sim 292^{\circ}.5$, the pattern stops and the lower density of stars in this domain appear to be closer, most likely in the foreground of an unsampled far arm -- if indeed the arm is continuous at these longitudes.  A final notable feature of the $\mu_{\ell,*}$ distribution is the island of data points near $\ell = 286^{\circ}$, at less negative values (near -5.0 mas yr$^{-1}$).  This is considered next.

\begin{figure}
\begin{center}
\includegraphics[width=1.0\columnwidth]{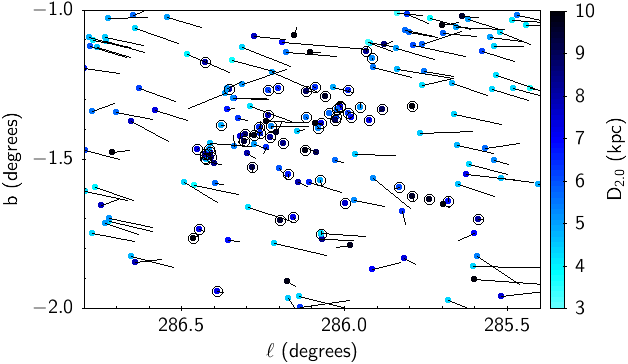}
\caption{
A map of the region in which the group of stars identified in Fig.~\ref{fig:pmsvlong} at less negative $\mu_{\ell,\ast}$ are found.  Relative proper motion vectors are attached to every star and are drawn on a scale with an offset origin that cancels the median proper motion ($\mu_{\ell,\ast} = -5.109$, $\mu_b = -0.172$ mas yr$^{-1}$) of the potential background association.  Possible members of the OB complex are shown encircled.  All are in the frame. By design, their proper motion vectors almost vanish. Relatively, the foreground stars stream preferentially toward lower longitudes and more negative latitude.  The stars of the group mainly locate in a dense elongated clustering running from $\ell \sim 285^{\circ}.9$ to $\ell \sim 286^{\circ}.4$.  The data points are coloured according to inferred distance, $D_{2.0}$, emonstrating that this grouping is most likely 7--8 kpc away. 
}   
\label{fig:group}
\end{center}
\end{figure}

\subsection{Evidence for a remote OB complex}
\label{sec:other}

The island of data points in the $\mu_{\ell,*}$ distribution at $\ell \simeq 286^{\circ}$ (Fig.~\ref{fig:pmsvlong}) locates on the sky roughly at $b = -1^{\circ}.4$.  There is no known cluster or association here.  To better isolate these stars, we have made a selection in the $\mu_{\ell,*}, \mu_b$ distribution, imposing a limit of 0.2 mas yr$^{-1}$ on $|\mu_{rel}|$, followed by a cut on sky position to identify the main over-density.  This gives an initial list of 58 objects, in which the median proper motion is $\mu_{\ell,*} = -5.109$, $\mu_b = -0.172$ mas yr$^{-1}$.  Their median Galactic coordinates are $(286^{\circ}.12, -1^{\circ}.40)$. The inferred median distance is $D_{2.0} = 7.36$ kpc, which is well beyond $D_{2.0} = 5.23$ kpc obtained for a $\Delta \ell = 1^{\circ}$ slice through the sample centered on the same longitude (299 stars). 
If the median components of proper motion of the group are used to estimate distance, and we assume their average motion is pure Galactic rotation, values of $\sim9$ kpc are obtained.  It remains to be seen, given the distance uncertainties, if this distant OB complex makes sense as an outrider to the Carina far arm or whether it is a sign of a distinct more distant structure. If the right interpretation of the Galactic longitude distribution of distance shown in Fig.~\ref{fig:pmsvlong} is that there is a break in the run of the far arm (see section~\ref{sec:structure1}), this group possibly represents the tip of the more distant segment.

Fig.~\ref{fig:group} shows the sky area including the 58 selected stars, along with proper motions plotted relative to the group median.  This all but cancels their motions and shows the many foreground stars streaming past, thanks to their more negative longitude motion.  There are examples of stars in Fig.~\ref{fig:group} that have small relative proper motion, and greater inferred distance not yet ringed as group members.  These may belong, given the present selection is more than preliminary.  It can be seen in the plot that many of the distant group lie in a strip of almost constant Galactic latitude, spanning a longitude range of $\sim 0.7$ degrees (a physical length of $\sim 90$ pc at 7.4 kpc).  Eleven of the 58 stars picked out have estimated effective temperatures consistent with O star status. Only one is flagged as a likely emission line star.  Most of the stars have $K$ magnitudes between 12 and 14, placing them in the fainter half of the range for the sample as a whole.  The brightest, \#1$-$03612, with $K = 10.108$ is also unusually red for the overall sample and may turn out to be a cool bright (or super-) giant. 


\section{The frequency of high relative proper motion massive stars}
\label{sec:PMstructure}

We have seen that the dispersion in the far arm is roughly $\pm0.6$ mas yr$^{-1}$ in $\mu_{\ell,\ast}$, and $\pm0.3$ mas yr$^{-1}$ in $\mu_b$.  At the sample median distance of 5.84 kpc, these translate to velocity dispersions in the plane of the sky of 17 km s$^{-1}$ and 9 km s$^{-1}$ respectively.  This means creates an opportunity  to identify outliers and examine options for a census of candidate runaway OB stars, mindful of the conventional threshold of a velocity contrast of 30 km s$^{-1}$ \citep[see e.g.][]{Hoogerwerf2001}.  First we attack this via the proper motions alone without regard for distance, aiming to identifier clear outliers.  Second, we fold in the distance information in order to apply cuts directly on tangential velocity, even if these will carry significant error thanks to the distance uncertainties.  In both approaches, the goal is a statistical appraisal of the relative proportions of O, B and emission-line stars qualifying as outliers, rather than a detailed list of identified objects.

\subsection{The vector point diagram and the selection of high proper motion stars}
\label{sec:hivel-select}

\begin{figure}
\begin{center}
\includegraphics[trim=7 15 25 25,clip,width=1.0\columnwidth]{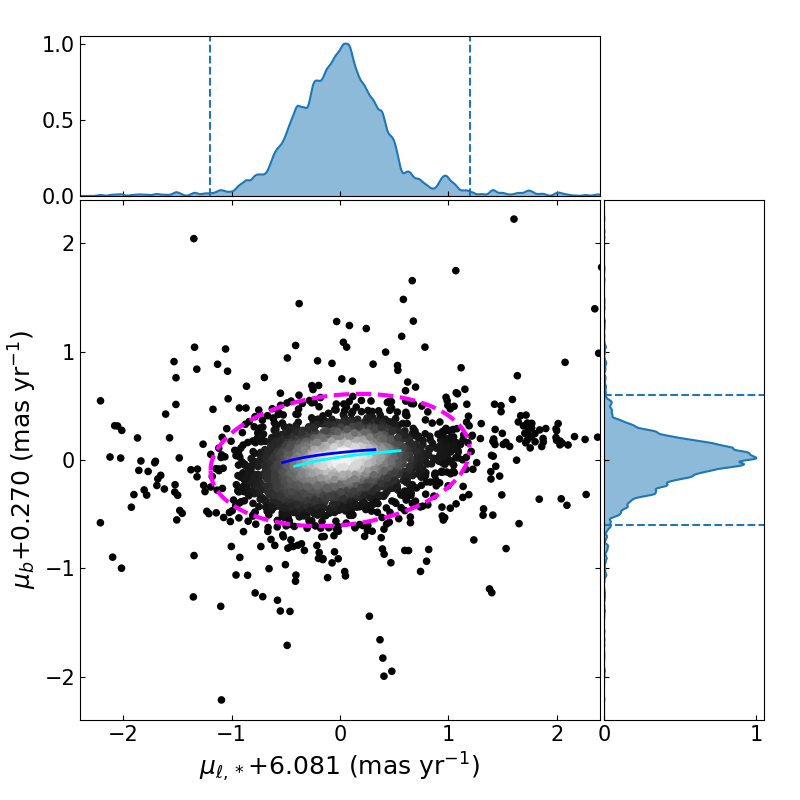}
\caption{
The proper motion vector point diagram of the selected OB stars as a density plot (greyscale with light for high density).  On each axis the overall sample median is subtracted in order to recentre the motions on (0,0).  The superposed cyan and blue line segments have been computed for $\ell = 285^{\circ}$, $4.5 < D < 7$ kpc and $\ell = 291^{\circ}$, $5 < D < 8$ kpc, respectively, in the case of pure Galactic rotation.  The magenta-dashed ellipse is the approximately $2\sigma$ boundary adopted between the core and outlier distributions. Its semi-major and semi-minor axes are 1.2 and 0.6 mas yr$^{-1}$.  The top and right-hand side panels provide the separate distributions of the recentered longitudinal (red) and latitudinal (blue) PM components,  normalised to their peaks.  They are KDE plots with Epanechnikov kernels, smoothing to 0.042 mas yr$^{-1}$).  The vertical lines mark the $\pm 1.2$ and $\pm 0.6$ mas yr$^{-1}$ semi-axes of the magenta ellipse in the main panel.
}   
\label{fig:pmlvpmb}
\end{center}
\end{figure}

We determine the distribution medians of the proper motion components to be $\mu_{\ell,\ast} = -6.081$ and $\mu_b = - 0.270$ mas yr$^{-1}$.  Hence the relative proper motion, ${\bf \mu_{{\rm rel}}}$, of any one star is $(\mu_{\ell,\ast} + 6.081$, $\mu_b + 0.270)$ in mas yr$^{-1}$.   
The vector point diagram (Fig.~\ref{fig:pmlvpmb} main panel) demonstrates the relative motions and confirms that the majority of the OB sample falls within a relatively compact distribution.  To provide a crude comparison and reference, we have calculated, for two Galactic longitudes and a range of heliocentric distances at each, the proper motions that would arise from pure circular rotation.  The results of this are plotted as the cyan and blue line segments: they clearly mimic the general trend seen in the data and show that the elongation of the central concentration is in part to do with the distance spread within the sample.  This will be further exaggerated by the modulation of $\mu_{\ell,*}$ with $\ell$ that is visible in the bottom panel of Fig.~\ref{fig:pmsvlong}. 

The outer panels of Fig.~\ref{fig:pmlvpmb} show the separate distributions of the two relative PM components, normalised to their peak values in order to bring out the difference in width. 
In the positive wing of the recentered $\mu_{\ell,*}$ distribution, a distinct group of stars introduces a subsidiary peak. These objects are from the island of stars at less negative $\mu_{\ell,*}$ (lowest panel of Fig.~\ref{fig:pmsvlong}) around $\ell \simeq 286^{\circ}$, that were discussed in section~\ref{sec:other}.

In order to review the incidence of high velocity stars in the sample, a criterion for separating off PM outliers is needed.  Preferring low contamination over high completeness, we have drawn a large ellipse around the main concentration in order to separate off just the far wings of the distribution. The ellipse chosen appears in the main panel of Fig.~\ref{fig:pmlvpmb} as a magenta dashed line. It has axes of 1.2 and 0.6 mas yr$^{-1}$ ($\sim2\sigma$ in each component), and it has been rotated by 6.15 degrees to line up with the small tilt visible in the vector point diagram. At the shorter $\sim4$ kpc distances for the sample (Fig.~\ref{fig:distance_hist}), these cuts are equivalent to cutting the longitudunal and latitudinal velocity components at $\sim$23 and $\sim11$ km s$^{-1}$.  These line up quite closely with the analogous cuts of 19 and 11 km s$^{-1}$ recommended by \cite{tetzlaff11}.  With increasing distance working through the OB sample the selection will become progressively more conservative than this, such that the on-average more distant O stars will be cut at higher tangential velocity than the B stars. The number of stars lying outside the elliptical boundary is 320, or 8\% of the far-arm sample. 

Even after our conservative selection, there remains an issue of clumps of outliers in Fig.~\ref{fig:pmlvpmb} lying to the right of the main locus of points.  They are positioned around $(1.8,0.2)$ in the coordinates of the plot: they, like the more prominent clustering pointed out in Section~\ref{sec:other}, could be stars lying behind the main run of the Carina far arm.  Indeed, when they are identified and inspected, they turn out to have $D_{2.0} > 8$ kpc and strongly favour $b \sim -2^{\circ}$. Altogether there are 32 of these objects, and we take them out of the outlier group. This leaves 288 stars.  

We can now ask how the fraction of O stars (photometric SED fits giving $\log T_{{\rm eff}} > 4.45$) in the outlier group compares with the fractions for the cooler early B stars, and the emission-line stars (both O and B type).  Out of the 4199 stars making up the sample, the total numbers in the O, B and emission-line categories are respectively 837, 2983 and 379.  When the common proper motion groups just noted are removed, the numbers adjust down to 824, 2915 and 370 (4109 stars altogether).

The outlier group of 288 objects breaks down into 83 O stars, 167 early B stars and 38 emission line objects. On forming the ratio with the relevant total for the trimmed OB-star sample, these numbers yield percentages of $(10.1 \pm 1.1)$\% and $(5.7 \pm 0.4)$\%, for non-emission O and B stars respectively (Poisson errors).  In other words, the O stars are almost twice as likely to exhibit high relative proper motion as the early B stars.  But in absolute terms, over twice as many early B stars, as compared to O stars, qualify as high relative PM stars.  Proportionately, the emission line stars are well represented in the outlier group in that $(10.3 \pm 1.7)$\% of them are included -- to be compared with $\sim$7\% for all their non-emission counterparts.  These outcomes are summarised in part (i) of Table~\ref{tab:vtrel_stats}.

\subsection{Selection according to relative tangential velocity}
\label{sec:vt_selection}

\begin{figure}
    \centering
    \includegraphics[width=1.0\columnwidth]{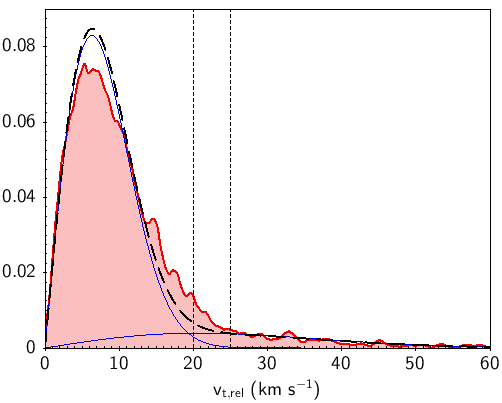}
    \caption{Area-normalised KDE distribution of relative tangential velocities for the trimmed Carina far-arm sample -- shown in red. Even in the presence of both significant distance uncertainties and some organised large-scale kinematic structure (Fig. 5, lower panel), this diagram retains the character of two overlapping 2D Maxwellians (blue lines).  Once $v_{t,rel}$ exceeds 25 km s$^{-1}$, all stars belong in the higher-velocity group.  Between 20 and 25 km s$^{-1}$ (marked as dashed vertical lines), it is evident that up to half the stars ($\sim$65 objects) are better viewed, statistically, as low velocity contaminants. This translates to an expectation that selecting for $v_{t,rel} > 20$ km s$^{-1}$ suffers $\sim$15\% contamination.  Requiring $v_{t,rel} >$ 25 or $>$30 km s$^{-1}$ will yield relatively clean samples.  
    }
    \label{fig:vtrel_hist}
\end{figure}

With the EDR3-inferred distance available for every OB star in the sample, it is straightforward to convert the relative proper motion into a relative tangential velocity, $v_{t,rel}$ for each one.  The difficulty with this comes from the large uncertainties in the distances, combined with the risk that some stars may lie significantly behind the far arm, beyond the bulk of the sample defining the median proper motion.  In contrast, distance precision is very much better on the near side of the sample ($D_{2.0} \lesssim 4$ kpc).  As mitigation for this inevitable asymmetry, we limit our attention in this exercise to the stars contained within the box drawn as a grey dashed line in the top panel of Fig.~\ref{fig:pmsvlong}: this is a selection in both $\ell$ and $D_{2.0}$ that aims to delimit the bulk of the far arm.  None of the stars in the distant common proper motion group discussed in section~\ref{sec:other} are included.  These steps reduce the sample to 3549 stars, for which the median proper motion is $\mu_{\ell,*} = -6.097$ mas yr$^{-1}$ and $\mu_b = -0.270$ mas yr$^{-1}$.

Multiplying the magnitude, $|\mu_{rel}|$, of the proper motions relative to the median (in mas yr$^{-1}$), by $4.74D_{2.0}$ kpc gives $v_{t,rel}$ for each star in km s$^{-1}$.  The error in this quantity is heavily dominated by the distance error. Its median value per individual object is close to 2 km s$^{-1}$, with the distribution presenting a long tail such that the 90$^{th}$ percentile is $\sim$ 8 km s$^{-1}$.  Higher error inevitably trends with increasing distance.  Usefully, for present purpose, the growth in error with increasing $v_{t,rel}$ is weak.  

To the relative tangential velocities, we apply 3 minimum-threshold cuts, and collect the statistics for each of the 3 subsets created.  The lowest cut applied ($v_{t,rel} > 20$ km s$^{-1}$) makes sense in terms of the approximate decomposition of the $v_{t,rel}$  distribution into two 2D Maxwellian distributions: these correspond to, respectively, the expected lower-velocity equilibrium scatter of the OB stars and the high-velocity candidate runaway component \citep[borrowing the technique exploited by][]{stone79,tetzlaff11}. The comparison is shown in Fig.~\ref{fig:vtrel_hist}. The cuts we have adopted are $v_{t,rel} > 20$, $> 25$, $> 30$ km s$^{-1}$. Only the lowest cut appears to suffer from significant contamination: on the basis of the data in Fig.~\ref{fig:vtrel_hist} we deduce this is most likely in the region of 15\%.  Like \cite{tetzlaff11}, we find that the cross-over point between the Maxwellian's describing the low-velocity stars and the candidate runaways is at $v_{t,rel} \simeq 20$ km s$^{-1}$.  

The proportions and numbers by object type of stars exceeding the three $v_{t,rel}$ thresholds are gathered in Table~\ref{tab:vtrel_stats}.  The proportions for $v_{t,rel} > 20$ km s$^{-1}$ are given with and without uniform rescaling for 15\% contamination.  The patterns, comparing O to B to the emission line group, closely parallel that obtained from the proper-motion selected sample in section \ref{sec:hivel-select}.  For  $v_{t,rel} > 30$ and $>25$ km s$^{-1}$, the non-emission O star fraction in the high-velocity tail is respectively 2.0 and 1.9 times the equivalent for non-emission B stars.  This drops to 1.6 for $v_{t,rel} > 20$ km s$^{-1}$ -- a change that may only reflect greater contamination.   It also remains the case that emission line stars are at least as likely as O stars to be runaway candidates (the fractions obtained are the same to within the Poisson errors).  The emission line stars are a mix of mainly Be stars with a smaller number of Oe and already-known WR stars.  

Stars that pass even the $v_{t,rel} > 30$ km s$^{-1}$ threshold do not necessarily also belong in the outlier group identified directly from relative proper motion via Fig.~\ref{fig:pmlvpmb}.  We find that 78\% of those above the 30 km s$^{-1}$ threshold are also proper motion outliers as defined by the method illustrated in Fig.~\ref{fig:pmlvpmb}. This drops to 51\% of the stars exceeding the 20 km s$^{-1}$ threshold (at the relevant sample median $D_{2.0}$ this threshold converts to $|\mu_{rel}| = 0.74$ mas yr$^{-1}$).  Nevertheless the most extreme objects, that fall outside the main plot frame in Fig.~\ref{fig:pmlvpmb}  easily meet both sets of criteria.  

\begin{table}
\caption{Percentages and numbers of O, B and emission line stars selected for (i) outlying relative proper motion, $|\mu_{rel}|$ fronm 4109 stars, or for (ii) exceeding relative tangential velocity ($v_{t,rel}$) thresholds out of 3549 stars. In both cases some trimming of the original sample of 4199 stars has been applied to deal with stars that may not belong to the far Carina arm (see text).  Under each star type, N, the relevant total number of stars of that type in the core sample is given, while n is the number per type selected. The errors on the percentages are Poisson errors. The second row of percentages, given in brackets for $v_{t,rel} > 20$ km s$^{-1}$, have been reduced to allow for a potential contamination of 15\%.   
}
{\centering
\begin{tabular}{l|cccccc}
\hline
  & \multicolumn{6}{c}{Star type} \\
  & \multicolumn{2}{c}{O} & \multicolumn{2}{c}{B} & \multicolumn{2}{c}{emission} \\
\hline
\multicolumn{7}{l}{(i) Relative proper motion ($|\mu_{rel}|$) selection} \\
   & \multicolumn{2}{c}{N = 824} & \multicolumn{2}{c}{N = 2915} & \multicolumn{2}{c}{N = 370} \\
   & \% & n & \% & n & \% & n \\   
\hline
  & & & & & & \\
outliers   &  10.1$\pm$1.1 & 83 & 5.7$\pm$0.4 & 167 & 10.3$\pm$1.7 & 38 \\
 & & & & & & \\  
\hline
\multicolumn{7}{l}{(ii) Relative tangential motion ($v_{t,rel}$) selection} \\
  & \multicolumn{2}{c}{N = 733} & \multicolumn{2}{c}{N = 2512} & \multicolumn{2}{c}{N = 304} \\ 
   & \% & n & \% & n & \% & n \\
\hline
$v_{t,rel}$ & & & & & & \\
km s$^{-1}$ & & & & & & \\
 $>$30 & 7.5$\pm$1.0 & 55 & 3.8$\pm$0.4 & 96 & 7.9$\pm$1.6 & 24 \\
 & & & & & & \\
 $>$25  & 10.2$\pm$1.2 & 75 & 5.4$\pm$0.5 & 136 & 10.9$\pm$1.9 & 33 \\
  & & & & & & \\
 $>$20  & 13.9$\pm$1.4 & 102 & 8.7$\pm$0.6 & 218 & 18.8$\pm$2.5 & 57 \\
   & (11.8$\pm$1.3) & & (7.4$\pm$0.5) & & (16.4$\pm$2.4) & \\
  & & & & & & \\
 \hline
\end{tabular}
 }
\label{tab:vtrel_stats} 
\end{table}

\subsection{The most extreme relative proper motion objects}
\label{sec:extreme}

\begin{table*}
\caption{Data on the objects with extreme relative proper motions ($|\mu_{rel}|$) and the highest relative tangential velocities ($v_{t,rel} > 100$~km s$^{-1}$).  Only stars with good parallaxes ($\pi/\sigma > 5$) and trustworthy SED fits ($\chi^2 < 7.82$) are included. 
The cited bounds on $D_{2.0}$ correspond to the 16th and 84th percentiles of the EDSD inference.  The lower bound is used to compute the lower limit on $v_{t,rel}$ given in brackets.  
}
{\centering
\begin{tabular}{rcccccrrrr}
\hline
 Cat. \# & \multicolumn{2}{c}{Galactic co-ordinates} &  $\log(T_{{\rm eff}} {\rm K})$ & $\chi^2$ & Parallax, $\pi$ & $\pi/\sigma$ & distance (min$-$max) & $|\mu_{{\rm rel}}|$ &  $v_{t,rel}$ (min)\\
  & $\ell^{\circ}$ & $b^{\circ}$ &  &  & mas &  & kpc &  mas yr$^{-1}$ & km s$^{-1}$ \\
\hline
 1$-$04845 &    286.97330 & 0.13208 & 4.379 & 0.48 & 0.14330 & 6.61 & 5.70 (4.9--7.4) & 13.94$\pm$0.02 & 377 (321) \\  
 1$-$04364 &    286.73827 & -0.84538 & 4.306 & 2.71 & 0.19787 & 8.15 & 4.38 (3.8--5.5) & 15.25$\pm$0.02 & 317 (274) \\ 
 1$-$03794 &    286.36969 & -2.96865 & 4.402 & 0.37 & 0.17978 & 14.22 & 4.76 (4.4--5.3) & 9.64$\pm$0.02 & 218 (199) \\ 
 1$-$00543 &    283.30684 & -0.80445 &  4.476 & 5.03 & 0.16626 & 6.82 & 5.05 (4.3--6.6) & 5.78$\pm$0.03 & 138 (118) \\
 2$-$15427 &    293.58643 & 0.63985  & 4.565 & 0.55 & 0.09258 &  7.78 & 8.01 (7.0--9.7)  & 3.00$\pm$0.01 &  114 (100) \\
\hline
\end{tabular}
 }
\label{tab:extreme} 
\end{table*} 

We comment, briefly, on the 5 most extreme objects in the sample with relative tangential speeds exceeding 100 km s$^{-1}$, for which robust measurements are available (from both MS17 and Gaia EDR3).  Key parameters for these stars are set out in Table~\ref{tab:extreme}.  The last quantity listed is the relative tangential speed, $v_{t_rel}$, computed using the distance inferred from the EDR3 parallax, and the magnitude of the relative proper motion.  The value in brackets is $v_{t,min}$, the minimum obtained on using the lower bound to the distance (also given in brackets in column 8 of the table). 

Two of the five stars have estimated effective temperatures high enough that they qualify as likely O-type (\#1$-$00543 and \#2$-$15427).  The second of these, and the last entry in Table~\ref{tab:extreme} is the only one not to satisfy the Galactic longitude constraint set for inclusion in the sample tested for $v_{t,rel}$. Its measured proper motion and inferred $D_{2.0}$ are nevertheless high enough to yield a tangential velocity in excess of 100 km s$^{-1}$. Three probable early B stars head up the list, moving across the sky at velocities relative to the far Carina arm of $\sim$200 km~s$^{-1}$ or more.  

The first two in the list, \#1$-$04845 and \#1$-$04364, are potentially runaways from Westerlund 2 that will be mentioned again in Section~\ref{sec:hotspots}. They set themselves well apart, even before correction to the Carina far arm frame, as the stars having by far the largest positive Galactic-longitude components of proper motion of any in the wider sample (respectively $+$7.70 and $+$8.80 mas yr$^{-1}$). The third B star, \#1$-$03794, only just falls inside the region near $b = -3^{\circ}$: it has a relatively high-quality parallax distance that places it firmly in among other far-arm stars at similar longitude.  

\section{Connecting runaway candidates to places of origin in the far-arm region}
\label{sec:runaways}

The outstanding massive young cluster in the studied region is NGC 3603.  An investigation of the pattern of O-star runaways around NGC 3603 was performed by \citet[hereafter D19]{Drew2019}, applying the assumption that ejections from the cluster would follow straight-line trajectories radiating away from cluster centre.  Since there can be no unique point of origin for all runaways in a cluster, an impact parameter criterion has to be set that requires every trajectory to pass within an upper limiting radius of cluster centre. In the case of NGC 3603, the chosen limiting radius of 1 arcmin enclosed the very compact core of the cluster \citep[radius of $\sim$4.8 arcsec, ][]{Harayama2008} and built in an allowance for impact parameter uncertainty.  A version of this quantitative method can now be applied across the wider far arm region and turned into a tool for creating a map of the potential origins of runaway stars.  We look at this next, beginning with an outline of the method applied and an assessment of the false positive rate, before moving onto the map of the ejection 'hot spots' it yields.  This sets the scene for revisitng NGC~3603 and Westerlund~2.  

\subsection{Method and wide-area simulations}
\label{sec:simulation}

The first step of the method is to choose a point of origin in the plane of the sky and the mean proper motion to associate with it to serve as reference.  For the case of NGC 3603, as outlined in D19, the mean proper motion was derived from 29 O stars within 1 arcmin of the adopted cluster centre.  With these two quantities in hand, the next step is to calculate the relative proper motion, ${\bf \mu_{{\rm rel}}}$, for each star in the available sample in order to identify those with magnitude, $|\mu_{{\rm rel}}|$ exceeding a preset threshold value (D19 used $> 0.6$ mas yr$^{-1}$, or $v_t > 20$ km s$^{-1}$ at 7~kpc). Then, for every star in the selection of higher $|\mu_{rel}|$ objects, the impact parameter, $p$, of the on-sky trajectory with respect to the reference position is determined from:  
\begin{equation}
    p = |{\bf r}|.\sin\theta
\end{equation}
where ${\bf r}$ is the position vector between the star and adopted centre in the plane of the sky, while $\theta$ is the angle between ${\bf \mu_{{\rm rel}}}$ and vector ${\bf r}$ (measured in radians).  

In the presence of error-free data, the upper limit on $p$ can be viewed as specifying a zone on sky, centered on the reference position, within which it would be plausible that ejections occur. In practice, this criterion also incorporates an allowance for error, acknowledging that selection is valid as long as the error in $p$ (propagated from $|{\bf r}|$ and $\theta$) is comparable with the applied upper bound on $p$.  Given the improvements in Gaia EDR3 relative to DR2, it is worth checking how the constraint, $p < 1$ arcmin, performs for the large sample now in hand.  Continuing to use NGC 3603 as the example, among the full list of 1163 stars across all 40 sq.deg, satisfying $|\mu_{{\rm rel}}| > 0.6$ mas yr$^{-1}$,
the median error in $p$ for the 110 stars within a radius of 1 degree is $\sim$0.5 arcmin.  The analogous Gaia DR2-based median errors were almost 3 times larger at $\sim 1.5$ arcmin.  The sky area considered by D19 was limited to $1.5\times1.5$ sq.deg within which the impact parameter errors would have remained compatible with the $p < 1$ arcmin constraint.  In view of the much larger sky area now under consideration and the reduction in proper motion errors, some testing of the search constraints is worthwhile.  Also of interest are the false positive rate, and likely biases or influences given the nature of the sample involved. A useful by-product will be a preliminary overview of likely ejection hot spots within the region.

\begin{figure}
\begin{center}
\includegraphics[width=0.95\columnwidth]{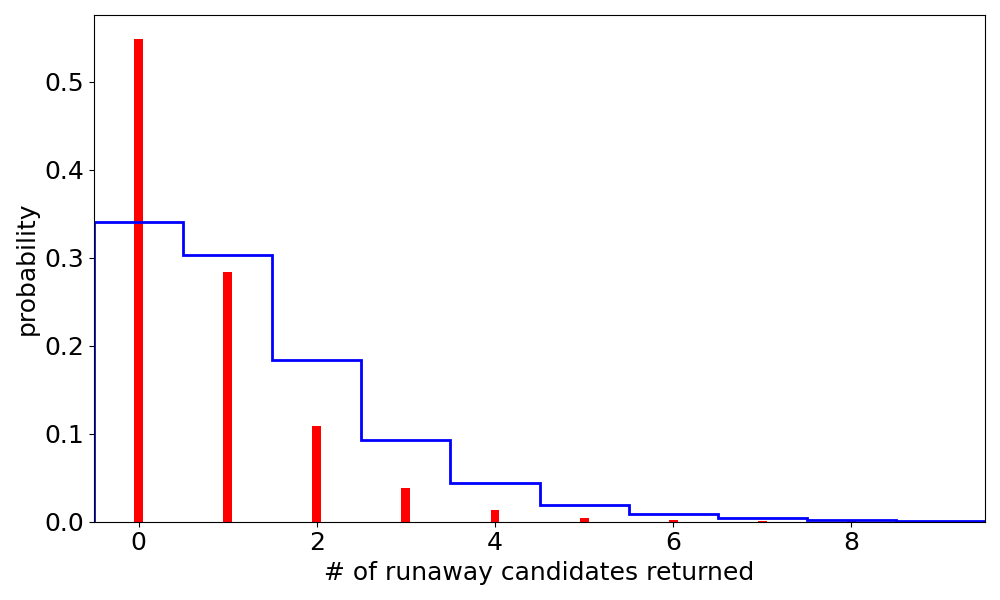}
\caption
{
Comparison of the histograms of candidate runaways identified in two simulations requiring $p < 1$ arcmin and (i) times of flight, $\Delta t < 2$ Myr (red spikes), (ii)  $\Delta t < 4$ Myr (blue steps).  Both simulations were run for a grid of 200000 sky positions, with random selection of the reference proper motion from around the overall sample median, at each position. In every trial, the individual-star proper motions were randomly reshuffled.  Both histograms are shown normalised to yield probabilities.  Over the first four bins, the distributions are close to Poisson in form. The respective means of the two distributions are 0.7 and 1.3 runaway candidates per simulated position and proper motion.  The 95th percentile candidate counts in the two cases are respectively 3 and 4 (higher than Poisson -- see text).
}
\label{fig:hist_candidates}
\end{center}
\end{figure}

\begin{figure*}
\begin{center}
\includegraphics[trim=-2 20 0 45,clip,width=2.1\columnwidth]{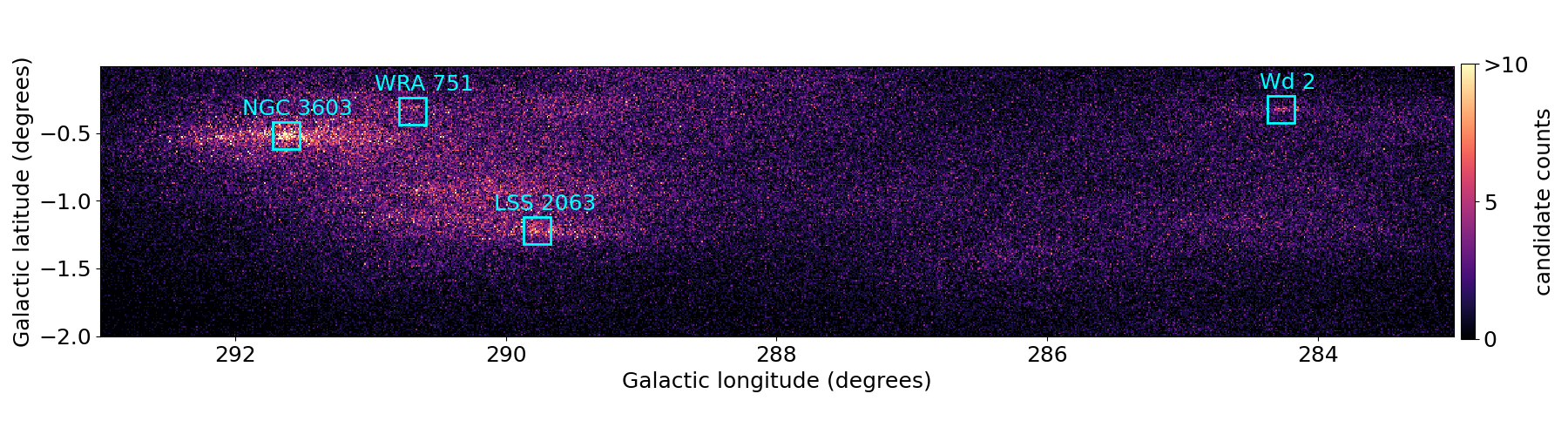}
\includegraphics[trim=0 12 0 45,clip,
width=2.1\columnwidth]{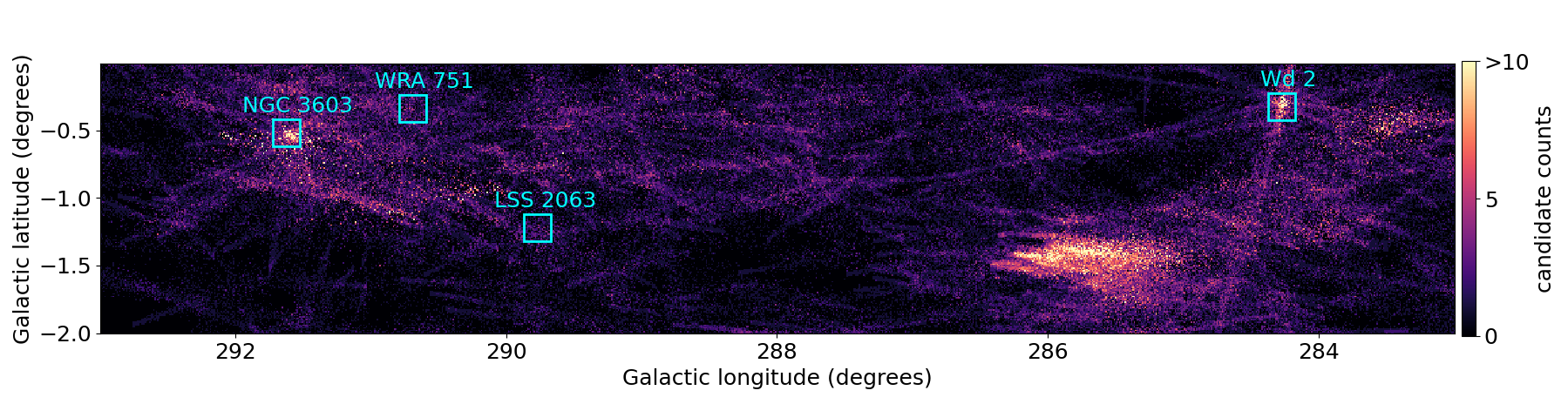}
\caption
{
Maps of the simulated numbers of runaway candidates per point of origin spanning a grid of 200000 positions within the central 20 sq.deg. region.  The constraints applied are $p < 1.0$ arcmin, $\Delta t < 4$ Myr, and $|\mu_{rel}| > 0.6$ mas yr$^{-1}$. The upper panel shows the map obtained when the proper motions measured for each of the 4199 stars in the sample are randomly reassigned to other star positions in each trial. The lower panel is the map that results when every star retains its measured proper motion throughout the simulation. The data points are coloured according to the runaway-candidate count obtained in the simulation.  The lower panel, using the observed pattern of proper motions, serves as a preliminary indication of the locations that may have been the seats of runaway-star production in the last 4 Myr.  The cyan open squares mark the positions of clusters as labelled in the lower panel: NGC 3603 and Westerlund 2 stand out as runaway 'hot spots'.  The diffuse bright patch offset from $\ell \sim 286^{\circ}$, $b \sim -1^{\circ}.5$ is linked with the common proper motion group described in Section~\ref{sec:other}.
}
\label{fig:hist_sim_maps}
\end{center}
\end{figure*}

In order to quantify these issues and understand the numerical effects at work, we have performed a series of wide-area simulations.  Every simulation is made up of 200\,000 trials across a uniform grid of positions sampling the test sky area, in which each trial (one per position) determines the number of stars with relative proper motion directed away from the grid point in question.  A reference proper motion is assigned to each grid point at the start of each trial, and then a pass is made through the data on all 4199 sample stars to compute the number of runaway candidates satisfying the imposed constraints on $p$ and $|\mu_{rel}|$.  From each simulation in which different $p$ and $|\mu_{rel}|$ constraints are chosen, a heat map of the test area can be constructed that shows the locations more or less likely to have been ejecting stars.  The data collected can also be presented in histogram form and analysed for the relevant statistics.

A further quantity collected in the simulations is the time of flight, $\Delta t$, for each candidate runaway from its point of origin.  We set an upper limit on it, and include it as a constraint alongside the constraints on $p$ and $|\mu_{rel}|$.  Querying for time of flight proxies in a more astronomically-relevant way for the angular range searched on sky.  The $\Delta t$ limits we favour are 2 and 4 Myr -- the former being a rough upper bound on the ages of NGC 3603, and also Westerlund 2. On sky and at the distance of NGC 3603, they roughly correspond to displacements of 0.5 and 1 degree, respectively (calculated for the median of the $|\mu_{{\rm rel}}| > 0.6$ objects: $\sim 0.82$ mas yr$^{-1}$).  For the less frequent higher velocity runaways the displacements will be proportionately larger.  As the limit on $\Delta t$ is raised beyond 4 Myr, typical errors on $p$ drift upwards from better than 1 arcmin, up to a median of $\sim2.2$~arcmin for $8 < \Delta t < 10$ Myr in our NGC 3603 example.  As this domain is approached, the ability to identify likely runaways is blurring and weakening the usefulness of the $p < 1$ arcmin constraint.  We take from this that the improved EDR3 astrometric errors permit credible results for flight times up to $\sim 5$ Myr.    

The grid of sky positions used in the simulations has a spacing 0.01 degrees in both $\ell$ and $b$.  The region examined is the central 20 sq.deg. of the full 40 sq.deg band (specifically:  $283^{\circ}.0 < \ell < 293^{\circ}.0$ and $-2^{\circ}.0 < b < 0^{\circ}.0$).  This means that no sky position queried is ever closer to the boundary of the overall OB sample than 1 degree, thereby reducing edge effects (for $\Delta t < 4$ Myr). 

The reference proper motion at each grid position is selected at random from normal distributions defined by $\overline{\mu_{\ell,\ast}} = -6.081$, $\sigma = 0.12$, and $\overline{\mu_b} = -0.270$, $\sigma = 0.06$.  The gaussian widths are set to enable sampling around the peaks of the distributions shown in the side panels of Fig.~\ref{fig:pmlvpmb}. This emulates the way in which we will search apparent runaway hot spots using locally-average reference proper motions that are only moderately displaced from the whole-sample $\mu_{\ell,*}$, $\mu_b$ medians.  Finally, in order to elucidate the false positive rate, the proper motions of the entire list of 4199 stars are randomly shuffled to different star positions, within each trial.  The idea here is to greatly weaken the relation that exists in the real data between the position of an OB star and the star's measured on-sky motion, whilst retaining the overall pattern of stellar density.   

The simulations we have run are summarised in Table~\ref{tab:stats}.  For every combination of constraint on $p$ and $|\mu_{rel}|$, outcomes were obtained for $\Delta t < 2$ and also $\Delta t < 4$ Myr: the effective increase in the sky area trawled shows up in the statistics (mean and 95th percentile) as increased candidate counts. In one instance, only, a simulation was run for $\Delta t < 8$ Myr to extend the trend.   The general pattern, as would be expected is that more permissive constraints lead to higher counts which, in these simulations, can be interpreted as the false positive rate.  We provide results for both $|\mu_{rel}| > 0.6$ and $>0.9$ mas yr$^{-1}$ aware that they map onto $v_{t,rel} \gtrsim 20$ km s$^{-1}$ at respectively the distances of NGC 3603 and Westerlund 2.

Fig.~\ref{fig:hist_candidates} compares the histograms of candidate runaway counts for the first two simulations listed in Table~\ref{tab:stats}. The distributions obtained are not far from Poissonian in the first four or five bins. 
The typical false-positive rate apparent from the figure is between 0 and 2, consistent with a Poisson mean not far from one (the actual distribution means are respectively 0.7 and 1.3).   However, the candidate-count high end tails seen in Fig.~\ref{fig:hist_candidates} are both elevated compared to the pure Poisson case: for means of 0.7 and 1.3 the expected 95$^{th}$ percentiles would be 2 and 3 -- to be compared with the values of 3 and 4 obtained from the simulations.  

\begin{table}
\caption{Statistics from simulation on how the number of chance candidates changes with changing simulation constraints.
}
{\centering
\begin{tabular}{ccc|cc}
\hline
\multicolumn{3}{c}{Constraints} & \multicolumn{2}{c}{Candidate count} \\
  $p$ & $|\mu_{rel}|$ & $\Delta t$ & mean & $P_{95\%}$ \\
  (arcmin) & (mas yr$^{-1}$) & (Myr) & & \\
\hline
\multicolumn{5}{c}{  } \\
\multicolumn{5}{l}{Sky region: $283^{\circ} < \ell < 293^{\circ}$, $-2^{\circ} < b < 0^{\circ}$} \\
$<1$ & $>0.6$ & $<2$ & 0.7 & 3 \\
     &        & $<4$ & 1.3 & 4 \\
     &        & $<8$ &  2.3 & 6 \\
     & $>0.9$ & $<2$ & 0.4 & 2 \\
     &        & $<4$ & 0.8 & 3 \\
$<0.5$ & $>0.6$ & $<2$ & 0.4 & 2 \\
     &          & $<4$ & 0.7 & 2 \\ 
\hline       
\multicolumn{5}{c}{  } \\
\multicolumn{5}{l}{Sky region: $289^{\circ} < \ell < 292^{\circ}$,$-1^{\circ}.4 < b < -0^{\circ}.2$:} \\ 
$<1$ & $>0.6$ & $<2$ & 1.5 & 4 \\
    &        & $<4$ & 2.6 & 6 \\
    &        & $<8$ & 3.9 & 8 \\
\hline
\end{tabular}
 }
\label{tab:stats} 
\end{table}

The understanding of where this comes from is aided by the map of the $10\times2$ sq.deg. simulation setting $dt < 4$~Myr, shown in the upper panel of Fig.~\ref{fig:hist_sim_maps}.  The obvious feature of the map is that the locations recording larger numbers of candidates occur mainly in the part of the Carina Arm running from around NGC 3603 down to the LSS 2063 association. Sampling this within the smaller region, $289^{\circ} \ell < 292^{\circ}$, $-1^{\circ}.4 < b < -0^{\circ}.2$ (bottom rows of Table~\ref{tab:stats}), we find the mean candidate count rises sharply to 1.5 and 2.6 for $dt < 2$ Myr and $dt < 4$ Myr (around twice the means obtained across the entire region).   On comparing the top panel of Fig.~\ref{fig:hist_sim_maps} with the OB-star density map in Fig.~\ref{fig:map}, it can be seen that there is a correlation between them.  Given the complete shuffling of the individual-star proper motions, the most likely explanation for this effect is that the higher density of sample stars in this part of the arm creates more opportunities for chance alignments.  It is of interest to note that there is only a hint of increase in false positives in the vicinity of Westerlund 2, where the overall stellar density is lower.  

Another property of the map is the streaking of high candidate counts at roughly constant Galactic latitude.  There is a particularly prominent streak passing through NGC 3603 that betrays the origin of this effect: it is the combined effect of clusterings (such as in NGC 3603), and the dominance of longitudinal proper motion in the OB-star sample.


The lower panel of Fig.~\ref{fig:hist_sim_maps}, computed for the same constraints as the upper panel shows the outcome of a final simulation in which every sample star retains its measured proper motion.  This map serves as an initial guide to where there can have been hot spots of runaway-star production over the past 4 Myr.  In this case, Westerlund 2 emerges clearly as a likely point of origin (in sharp contrast to the upper panel of the figure). At the higher longitudes, NGC 3603 is also prominent, but it is not alone in that the region showing high counts in the upper panel of Fig.~\ref{fig:hist_sim_maps} is again present.  Both clusters present combinations of position and reference proper motion that yield upward of 15 candidate ejections (whilst the colour scales in Fig.~\ref{fig:hist_sim_maps} saturate at 10).  In clear contrast to this, neither the WRA 751 cluster nor the LSS 2063 association stand out from the background false positive rate of 4 in their localities. If either clustering has ejected stars in the past, the implication of the simulation is that few, if any, ejections have taken place in the last 4 Myr.    

The smear of prominent hot spots near $\ell \sim 286^{\circ}$, $b \sim -1^{\circ}.4$ has a different origin in that it is created by the distant group of stars discussed in section~\ref{sec:other}.  This points out how this method of searching also exposes coherent kinematic behaviour such as that long recognised and exploited in moving groups.  Indeed, detailed inspection of the elevated candidate counts near NGC 3603 shows that the instances of high counts returned typically relate to groups of stars moving coherently. For some locations, the 'group' picked out is made up of stars in NGC 3603 itself.  So, before a source of runaway activity can be claimed, it is necessary to examine the candidate source region more closely and also consider the on-sky pattern of runaway candidates.  Where a cluster or moving group is implicated, the candidates will cluster, rather than exhibit a spread around the reference position.

\subsection{NGC 3603 and Westerlund 2, revisited}
\label{sec:hotspots}

Armed with a better idea of what to expect from a search, we re-examine NGC 3603 and Westerlund 2, that have emerged as the two outstanding hot spots in the region for the production of runaways (lower panel of Fig.~\ref{fig:hist_sim_maps}). 

\begin{figure}
\begin{center}
\includegraphics[width=1.0\columnwidth]{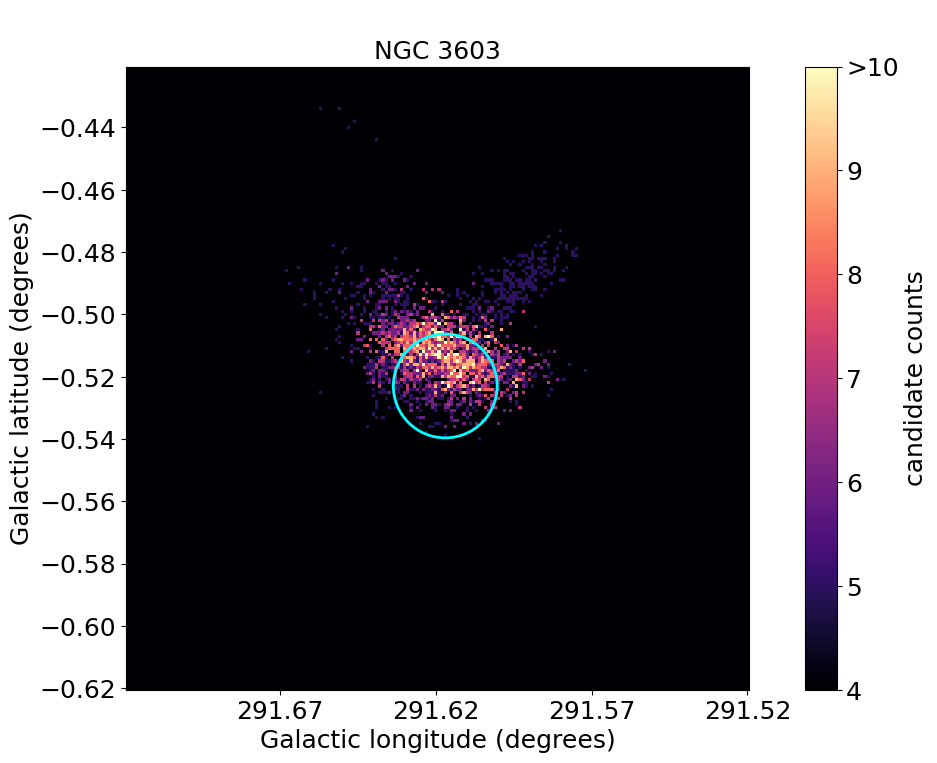}
\includegraphics[width=1.0\columnwidth]{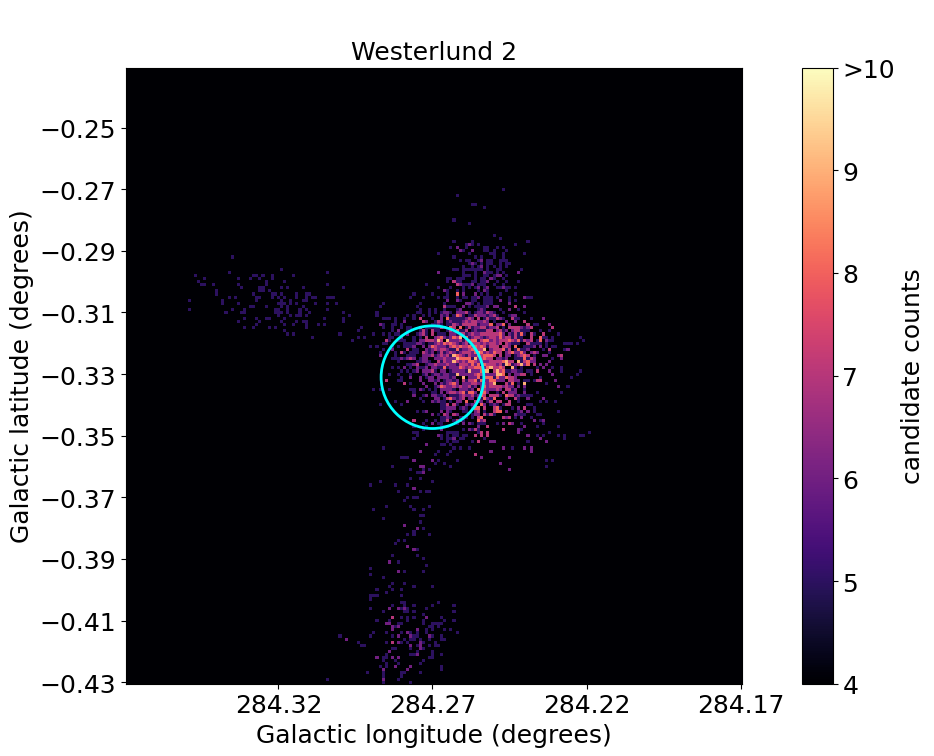}
\caption
{Maps of the returned numbers of candidate ejections per point of origin from simulations performed within $12\times12$ sq.arcmin. boxes centered on NGC 3603 (top) and Westerlund 2 (bottom).  In both cases we tighten the search limits to $p < 0.5$~arcmin and $\Delta t < 2$ Myr.  A difference is that, for Westerlund 2, the minimum required $|\mu_{rel}|$ is increased to 0.9 mas yr$^{-1}$ to compensate for the appreciably shorter distance to the cluster (4.4 kpc compared to $\sim 7$ kpc for NGC 3603).  The colour scale applied is the same in both panels and has been set such that all locations reporting 4 or fewer candidate ejections are dark. In both panels, a cyan 1 arcmin circle is drawn around the expected cluster centre position. In both cases, the preferred centre for candidate ejections is shifted a little away from nominal centre.
}
\label{fig:clusters_sim_1}
\end{center}
\end{figure}

\begin{figure*}
\begin{center}
\includegraphics[width=1.6\columnwidth]{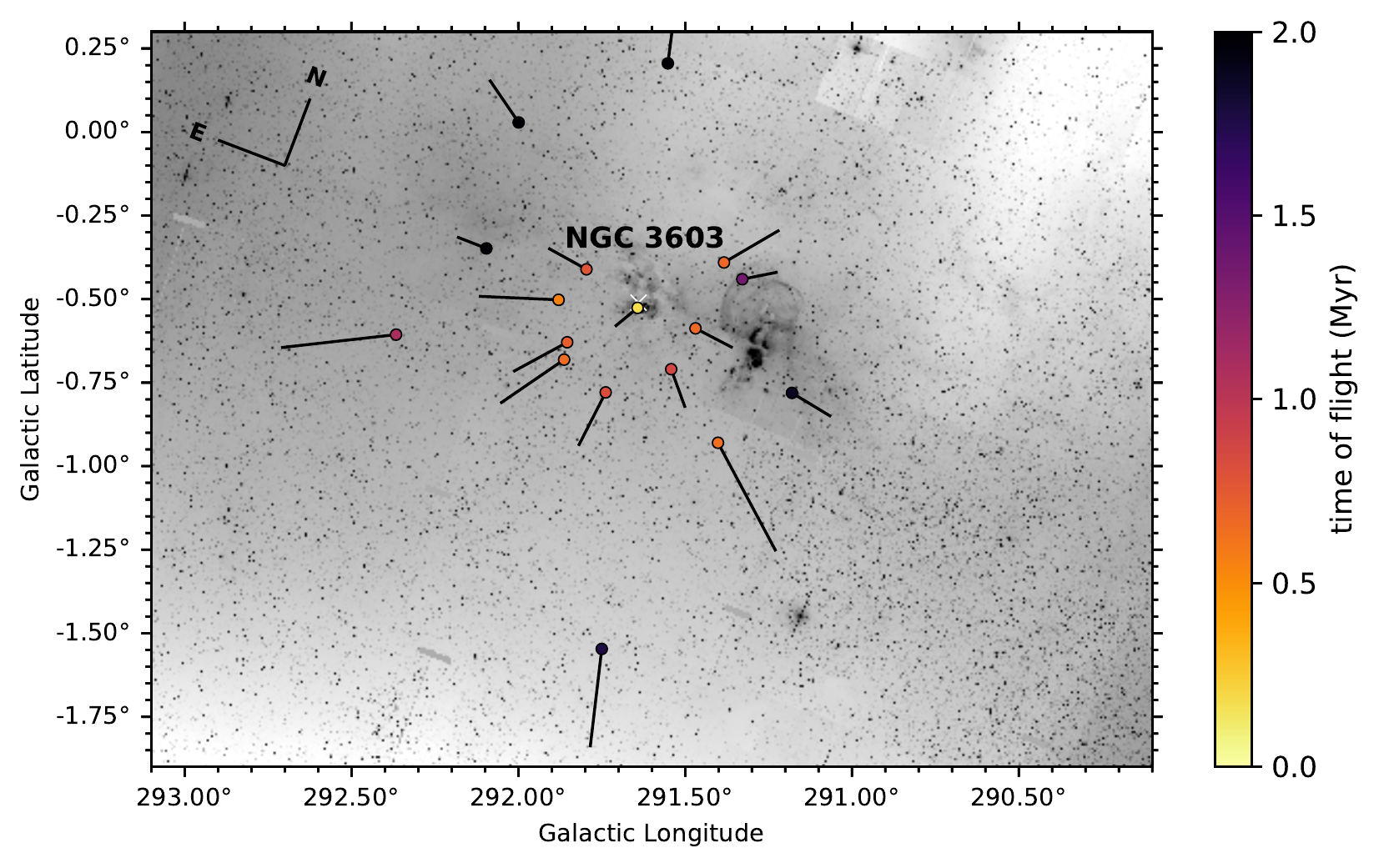}
\includegraphics[width=1.6\columnwidth]{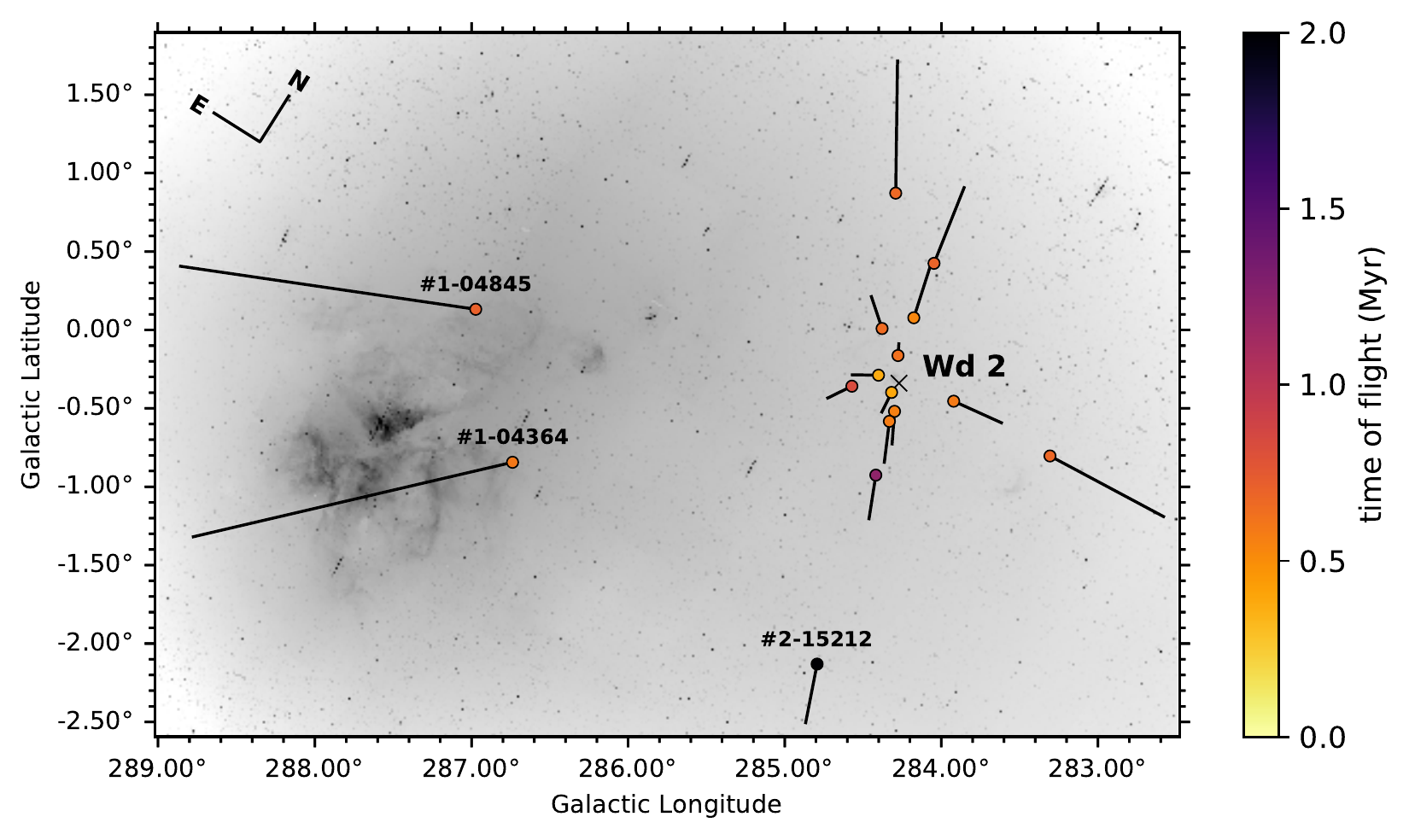}
\caption
{Sky maps of candidate ejections from NGC 3603 (top) and 
and Westerlund 2 (bottom).  The arrows drawn show the direction of relative proper motion and have lengths proportional to the magnitude of the motion, showing how far each star would travel in 0.5 Myr (see Table~\ref{tab:n3603_wd2}).  The included objects are represented by filled circles coloured according to time of flight.  The outstanding change relative to the findings of D18 and D19 is the pair of fast runaway B stars at higher $\ell$ relative to the centre of Westerlund 2.  The cluster centre of Westerlund 2 is marked by a black cross.  The centre of NGC 3603 is very close to the position of the star coloured yellow.
}
\label{fig:runaway_maps}
\end{center}
\end{figure*}

Simulations zooming into the neighbourhoods of the two clusters have been performed to better understand what is visible in the lower panel of Fig.~\ref{fig:hist_sim_maps}.  We can also use these more focused simulations to optimise the choice of reference position in the candidate search. To narrow the centering down, we tighten the impact parameter limit to $p < 0.5$ arcmin.
 Now, the reference proper motion is tied to the median cluster value.   There is still some randomisation but it draws from a much tighter normal distribution centered on the cluster median, identifying $\sigma$ with the dispersion (standard deviation) among the subset of stars defining the median.  The sky area explored is reduced to 12$\times$12 arcmin$^2$, and it is sampled using a grid of cell size 0.06$\times$0.06 arcmin$^2$ (40\,000 sky positions). 
 
 For each cluster, the median is constructed from the OB stars lying within 1 arcmin of a plausible first choice for the on-sky cluster centre.  In the case of NGC 3603, there are 34 OB stars within 1 arcmin of the position $\ell = 291^{\circ}.617, b = -0^{\circ}.523$.  The median proper motion of these stars is $\mu_{\ell,*} = -5.925\pm0.027$, $\mu_b = -0.147\pm0.019$ mas yr$^{-1}$.  For Westerlund 2, 26 OB stars lie within 1 arcmin of $\ell = 284^{\circ}.270$, $b = -0^{\circ}.331$.  These yield a median proper motion of $\mu_{\ell,*} = -6.117\pm0.038$, $\mu_b = -0.437\pm0.030$ mas yr$^{-1}$.


The maps resulting from these focused simulations are shown in Fig.~\ref{fig:clusters_sim_1}.
For both clusters the reference positions returning the highest numbers of runaway candidates fall reassuringly close to the initially-chosen cluster centres.
But in neither case is there perfect alignment, the offsets being of the order of an arcminute.  Small shifts like this are comparable with cluster dimensions, and may reflect one or more of the following: the limitations of assuming straight-line trajectories for the ejected objects; change over time of the clusters' median proper motion; remaining systematics in the astrometric data.  We note that the sky area within which the highest numbers of candidate ejections might have occurred is between 1 and 2 arcmin across. This is consistent with the effective cross-section defined in the simulation by the $p < 0.5$ arcmin criterion.  

\begin{table*}
\caption{Data on candidate OB-star ejections from (a) NGC 3603, (b) Westerlund 2 . In each section, they are ordered by estimated time of flight, $\Delta t$. 
For NGC 3603 ($D_{2.0} = 6.91$ kpc), the minimum relative proper motion $|\mu_{rel}|$ required of ejection candidates is 0.6~mas yr$^{-1}$.  This was increased to 0.9~mas yr$^{-1}$ for Westerlund 2 ($D_{2.0} = 4.41$ kpc), in view of its lower distance.   The last few rows at the foot of each section identify 'near miss' objects of potential interest (see text). The errors on $\Delta t$ include contributions from the uncertainty in $|\mu_{rel}|$ and a generous fixed error of 1.0~arcmin in $|{\bf r}|$ that acknowledges our ignorance of the exact point of ejection within either cluster. The final column indicates whether the object was already identified previously as a likely runaway by either D18 (Westerlund 2) or by D19 (NGC 3603). 
}
{\centering
\begin{tabular}{rccrcccccc}
\hline
 cat \# & Gaia EDR3 ID & $\ell$,$b$ & ${\bf |r|}$ & $p$ & $|\mu_{{\rm rel}}|$ & $\Delta t$ & $D_{2.0}$ (range) & type &  \\
  & & (degrees) & arcmin & arcmin & mas yr$^{-1}$ & Myr & kpc & & \\
\hline
\multicolumn{2}{l}{NGC 3603} & & & & \\
\hline
  1$-$13650 & 5337417813683621120 & 291.643,$-$0.526 & 1.9 & 0.26$\pm$0.17 & 0.63$\pm$0.06 &  0.18$\pm$0.10 & 9.35 (6.5$-$17.6)   & B  &   \\
  1$-$13931 & 5337403073354793088 & 291.880,$-$0.502 & 15.8 & 0.25$\pm$0.12 & 1.72$\pm$0.03 &  0.55$\pm$0.04 & 8.41 (6.9$-$11.8)  &  O  & Y \\
  1$-$13299 & 5337200247815920640 & 291.402,$-$0.930 & 28.3 & 0.57$\pm$0.39 & 2.65$\pm$0.05 &  0.64$\pm$0.03 & 5.74 (4.4$-$10.3) & B  &   \\  
  1$-$13918 & 5337027693199975040 & 291.863,$-$0.681 & 18.0 & 0.14$\pm$0.23 & 1.67$\pm$0.03 &  0.65$\pm$0.04 & 6.93 (5.9$-$9.1) & EM & Y \\
  1$-$13362 & 5337046316178212864 & 291.469,$-$0.588 & 10.0 & 0.09$\pm$0.15 & 0.90$\pm$0.03 &  0.66$\pm$0.07 & 6.55 (5.8$-$7.8)  & O & Y \\ 
  1$-$13280 & 5337612495964116096 & 291.384,$-$0.391 & 15.7 & 0.84$\pm$0.30 & 1.38$\pm$0.04 &  0.68$\pm$0.05 & 7.38 (5.9$-$11.4) & O & Y  \\  
  1$-$13908 & 5337403554391015424 & 291.854,$-$0.629 & 16.0 & 0.51$\pm$0.34 & 1.33$\pm$0.04 &  0.72$\pm$0.05 & 6.31 (5.2$-$9.0) & O & Y \\
  1$-$13860 & 5337412487923505792 & 291.796,$-$0.411 & 12.3 & 0.05$\pm$0.23 & 0.94$\pm$0.03 &  0.79$\pm$0.07 & 6.16 (5.3$-$8.0) & O & Y \\
  1$-$13804 & 5337032641002080640 & 291.739,$-$0.780 & 17.8 & 0.80$\pm$0.67 & 1.29$\pm$0.05 &  0.83$\pm$0.06 & 5.43 (4.2$-$9.8) & O & Y \\  
  1$-$13436 & 5337043739197624064 & 291.542,$-$0.710 & 12.8 & 0.05$\pm$0.20 & 0.88$\pm$0.03 &  0.87$\pm$0.07 & 7.37 (6.4$-$9.2) & O & Y \\ 
  1$-$14451 & 5337342806372428032 & 292.367,$-$0.607 & 45.4 & 0.72$\pm$1.13 & 2.50$\pm$0.06 &  1.09$\pm$0.04 & 5.78 (4.3$-$11.6) & B &   \\ 
  1$-$13230 & 5337611533891352064 & 291.329,$-$0.441 & 17.7 & 0.72$\pm$0.92 & 0.78$\pm$0.04 &  1.37$\pm$0.11 & 6.41 (5.1$-$10.4) & B &   \\
  1$-$13820 & 5336895717442641536 & 291.750,$-$1.548 & 62.8 & 0.71$\pm$0.77 & 2.13$\pm$0.03 &  1.77$\pm$0.04 & 8.00 (6.5$-$11.7) & B &   \\ 
  1$-$12992 & 5337240826669573504 & 291.180,$-$0.781 & 30.8 & 0.42$\pm$0.58 & 0.99$\pm$0.03 &  1.88$\pm$0.09 & 7.12 (6.0$-$9.5) & O &    \\
  1$-$14042 & 5337397678879696256 & 291.999,\ \ 0.029  & 39.7 & 0.75$\pm$1.74 & 1.11$\pm$0.06 &  2.14$\pm$0.13 & 4.99 (3.8$-$9.3) & B &    \\
\hline  
  2$-$15331 & 5240941990532685184 & 291.502,$-$2.584 & 124.7 & 1.14$\pm$2.16 & 3.29$\pm$0.05 & 2.28$\pm$0.04 & 5.74 (4.4$-$10.1) & B &    \\
  1$-$13452 & 5337557657842299008 & 291.552,\ \ 0.206 & 43.1 & 1.72$\pm$0.58 & 0.97$\pm$0.03 & 2.66$\pm$0.09 & 7.29 (6.2$-$9.5) & O & Y \\
  1$-$14172 & 5337385751715379840 & 292.096,$-$0.348 & 30.4 & 1.46$\pm$0.61 & 0.68$\pm$0.03 &  2.68$\pm$0.16 & 7.24 (6.1$-$9.8) & O &  Y \\
\hline
\multicolumn{2}{l}{Westerlund 2} & & & & & & \\
\hline
  1$-$01354 & 5351707101133943552 & 284.401,$-$0.289 & 9.2 & 0.02$\pm$0.12 & 1.50$\pm$0.05 & 0.37$\pm$0.04 & 4.73 (4.1$-$5.8)  & O &   \\
  1$-$01294 & 5255676167891788800 & 284.318,$-$0.399 & 6.2 & 0.48$\pm$0.20 & 0.95$\pm$0.04 & 0.39$\pm$0.07 & 4.17 (3.6$-$5.2) & O &   \\
  1$-$01119 & 5351766715283439104 & 284.176,\ \ 0.078 & 24.4 & 0.59$\pm$0.30 & 2.78$\pm$0.04 & 0.53$\pm$0.02 & 5.13 (4.2$-$7.8) & EM & Y \\
  1$-$01273 & 5255669399023171456 & 284.299,$-$0.520 & 12.3 & 0.00$\pm$0.14 & 1.31$\pm$0.03 & 0.56$\pm$0.05 & 4.52 (4.2$-$5.1) & O & Y \\ 
  1$-$01308 & 5255667681036173568 & 284.332,$-$0.584 & 16.5 & 0.72$\pm$0.15 & 1.72$\pm$0.03 & 0.58$\pm$0.04 & 4.62 (4.3$-$5.1) & EM & Y \\
  1$-$00972 & 5255691114377445376 & 283.921,$-$0.455 & 21.4 & 0.97$\pm$0.13 & 2.19$\pm$0.04 & 0.59$\pm$0.03 & 4.02 (3.7$-$4.4) & O &   \\
  1$-$04364 & 5254499479976619264 & 286.738,$-$0.845 & 152.5 & 0.07$\pm$0.24 & 15.25$\pm$0.04 & 0.60$\pm$0.01 & 4.38 (3.8$-$5.5) & B &   \\ 
  1$-$01236 & 5351757438154160000 & 284.276,$-$0.164 & 9.6 & 0.07$\pm$0.17 & 0.91$\pm$0.04 & 0.63$\pm$0.07 & 4.90 (4.4$-$5.7) & O & Y \\
  1$-$01338 & 5351717851422618496 & 284.378,\ \ 0.009 & 21.2 & 0.01$\pm$0.10 & 1.95$\pm$0.03 & 0.65$\pm$0.03  & 5.03 (4.6$-$5.6) & O & Y \\
  1$-$00543 & 5258728652676886400 & 283.307,$-$0.804 & 63.6 & 0.79$\pm$0.27 & 5.67$\pm$0.05 & 0.67$\pm$0.01 & 5.05 (4.3$-$6.6) & O &   \\
  1$-$04845 & 5350519766026431488 & 286.973,\ \ 0.132 & 165.7 & 0.28$\pm$0.23 & 14.00$\pm$0.04 & 0.71$\pm$0.01 & 5.70 (4.9$-$7.4) & B &   \\  
  1$-$01374 & 5255633871052073728 & 284.418,$-$0.927 & 37.7 & 0.45$\pm$0.35 & 1.86$\pm$0.03 & 1.22$\pm$0.03 & 4.91 (4.4$-$5.7) & O & Y \\
 2$-$15212 & 5255058139280372608 & 284.793,$-$2.134 & 111.5 & 0.22$\pm$1.39 & 2.58$\pm$0.04 & 2.64$\pm$0.04 & 3.85 (3.3$-$4.8) & B &   \\ 
\hline
  1$-$01260 & 5351855157229904768 & 284.291,\ \ 0.874 & 71.7 & 1.22$\pm$0.64 & 6.44$\pm$0.04 & 0.67$\pm$0.01 & 3.47 (3.1$-$4.2) & B &   \\
  1$-$01046 & 5351803514564210560 & 284.047,\ \ 0.426 & 46.4 & 1.86$\pm$0.15 & 4.02$\pm$0.03 & 0.69$\pm$0.02 & 4.33 (4.0$-$4.8)  & O & Y \\ 
  1$-$01485 & 5255625487274545664 & 284.571,$-$0.359 & 19.3 & 1.68$\pm$0.54 & 1.39$\pm$0.07 & 0.83$\pm$0.06  & 3.93 (3.1$-$6.8) & B &   \\
\hline
\\
\end{tabular}
 }
\label{tab:n3603_wd2} 
\end{table*}

The next step in the search is to choose an adjusted reference position for each cluster and then identify OB stars that meet the imposed upper limits on $p$, $\Delta t$, and lower limit on the magnitude of relative proper motion $|\mu_{rel}|$.  To produce a reasonably complete 'long list' of candidates to scrutinise, these are set generously. 

For NGC 3603, the finally adopted reference position is $\ell = 291^{\circ}.616$, $b = -0^{\circ}.510$, while the search limits set are $p < 2$ arcmin, $\Delta t < 4$ Myr, and $|\mu_{rel}| > 0.6$ mas yr$^{-1}$.  At the cluster distance of 7.27~kpc, computed in section~\ref{sec:structure1}, the minimum on $|\mu_{rel}|$ converts to a tangential velocity of 20.6 km s$^{-1}$.  The objects identified by the search are listed, along with selected parameters, in the top half of Table~\ref{tab:n3603_wd2}.  Fifteen candidates satisfying $p < 1$ arcmin are found, along with three more 'near misses' with impact parameters between 1 and 2 arcmin. 

As the selection technique applied here is much the same as applied by D19 to NGC 3603, the emergent list of candidate runaways from this cluster has much in common. Of eleven good or possible candidates presented by D19, eight reappear here in the main list.  The three near misses, including two stars from D19, listed at the foot of part (a) in Table~\ref{tab:n3603_wd2}, are classified as such on the grounds that $p > 1$ arcmin and $\Delta t > 2$ Myr (requiring ejection potentially before NGC 3603 formed). D19 already expressed caution regarding one of them -- the case of \#1$-$13452.  Finally, object \#1$-$13519 mentioned by D19 is absent from the new list because it is not in the main sample to start with (its revised EDR3 parallax places it plausibly in the foreground).  All the B type candidates ($\log(T_{{\rm eff}} < 4.45$) are new additions, since D19 considered candidate O stars only.  


For Westerlund 2, the finally adopted reference position is $\ell = 284^{\circ}.251$, $b = -0^{\circ}.321$.  The offset from the original choice of centre in this case is 1.3 arcmin (to be compared with 0.8 arcmin for NGC 3603).  The search limits set on $p$ and $\Delta t$ are the same as for NGC 3603. However, a different limit on $|\mu_{rel}|$ is appropriate given that the distance to this cluster is much less (4.44 kpc, see section~\ref{sec:structure1}).  As this is about 2/3 the distance computed in the same way for NGC 3603, the minimum on $|\mu_{rel}|$ is raised to 0.9 mas yr$^{-1}$ to bring it close to $v_{t,rel} = 20$ km s$^{-1}$ in the plane of the sky. The objects identified by the search performed on this basis are listed, along with selected parameters, in part (b) of Table~\ref{tab:n3603_wd2}.  Thirteen stars are found that satisfy $p < 1$ arcmin.

The overlap between the Westerlund 2 entries in Table~\ref{tab:n3603_wd2} with the list constructed by D18, is extensive as well: six of nine candidates are re-confirmed, with two more picked up as 'near misses'.  The objects dropping out altogether are \#1$-$01102 and \#1$-$01133 on account of what is now too low a magnitude of relative proper motion (respectively $|\mu_{rel}| = 0.76$ mas yr$^{-1}$ and 0.88 mas yr$^{-1}$). Of course it remains open that these stars are ejections -- should spectra of them reveal large enough radial velocities to permit them to qualify.  That there is overlap with respect to D18 is not a given, in view of clear differences in method and the shift of the reference position and the new Gaia data release.  The selection made by D18 rested on a more heterogeneous set of criteria in which the only astrometric element was a relatively loose constraint on the trajectory angle, $\theta$. Only likely O stars were investigated subject to the additional constraints of at least 10 arcmin separation from cluster centre and an interstellar extinction, $A_0 > 5.5$ (for compatibility with location in the Carina Arm tangent region). The simplified criteria used here have resulted in the addition of seven stars to the list: three B stars, two O stars inside 10 arcmin, and two newly-qualifying O stars thanks mainly to the shift in reference position.

Fig.~\ref{fig:runaway_maps} shows the pattern of higher-confidence runaways around NGC 3603 and Westerlund 2, using the data presented in Table~\ref{tab:n3603_wd2}.  For NGC 3603, the picture of a ring of runaways put forward by D19 is endorsed.  The main change is the uncovering of some more candidate runaways at larger on-sky angular separations.  These added objects are generally earlier ejections, with just one of them exceeding $\Delta t = 2.0$ Myr.  It remains the case that there seems to have been a rush of ejections between 0.5 and 1.0 Myrs ago (9 objects, mostly in the ring). The relevant data in Table~\ref{tab:stats} (third row from bottom) suggest that the number of false positives in this case -- given the favouring of $\Delta t < 2$ Myr -- is most likely to be around two out of the full list of 15 candidates. 
  
In the case of Westerlund 2, the tendency for the candidate ejected stars to lie along an axis almost perpendicular to the Galactic equatorial plane is diluted by the new additions. On the other hand, a feature that grows in prominence is the strong $\Delta t$ clustering: flight times are now heavily concentrated in the range $\sim$0.3 to $\sim$0.7 Myrs (11 of 13 stars).  Just as striking are the two newly-added B stars, \#1$-$04364 and \#1$-$04845.  These appeared already in Table~\ref{tab:extreme} on account of being the highest relative proper motion stars listed in the entire sample.  The inferred distance to \#1$-$04364, $D_2.0 = 4.38$ kpc, is a very good match to the Westerlund 2 distance, $D_2.0 = 4.41$ kpc.  For \#1$-$04845 it is a concern that the distance may be too great for compatibility ($D_{2.0} = 5.70$ kpc, with a lower bound of 4.9 kpc).  The time since ejection rises above 2 Myr just once, for the case of \#2$-$15212.  The expectation, based on the simulation summary statistics presented in Table~\ref{tab:stats}, 4th row, would be that at most two false positives may have arisen.  As a clear $\Delta t > 2$ Myr outlier (Table~\ref{tab:n3603_wd2}), perhaps \#2$-$15212 is one.   

The relative radial velocities of \#1$-$01273 and \#1$-$01338 were measured by D18.  They are respectively $-11$ and $-26$ km s$^{-1}$.  When combined with the EDR3 relative proper motions and our parallax-based estimate of the distance to Westerlund 2, the interquartile ranges of the relative space velocities of these two stars become 27--33 km s$^{-1}$ and 45--53 km s$^{-1}$.   This confirms and tightens the results reported by D18.

\subsection{Older clusters -- increasing the time of flight range searched}


Both NGC 3603 and Westerlund 2 are very young ($<2$ Myr).  In such cases there is no reason to expect a search for runaway candidates with times of flight well in excess of cluster age. There are older clusters for which it would make sense to try doubling the time of flight range considered.  One such would be the cluster associated with the luminous blue variable, WRA~751 \citep{vanGenderen01}: it has been argued by \cite{Pasquali06} that it is not less than 4 Myr old.  We have performed a runaway search for this cluster and for the LSS~2063 association (for which there is no age constraint presently), extending the time-of-flight range to 8 Myr.  In every other respect the method applied was the same as for Westerlund 2 and NGC 3603. 

In brief, the numbers of candidates returned were comparable with the expected false positive rates.  Both clusterings fall in the higher density part of the region to which the last rows in Table~\ref{tab:stats} apply.  The data there indicate a median false positive rate of 4 (with the potential to rise as high as 8).  For WRA 751, 8 candidates satisfying $p < 1$ arcmin were picked out, while for LSS 2063 the equivalent count was only 4. Loosening the impact parameter to $p < 2$ arcmin pulled in a few more, but we noticed that both candidate lists exhibited a bias toward peculiar motion within the Galactic plane ($|\mu_{\ell,*}| >> |\mu_b|$), along with no particular time-of-flight bias (almost as many in flight for over 4 Myr, as under).  We regard the exercise as inconclusive and as a warning that searching beyond $\Delta t = 4$ Myr may yield diminishing returns.

\section{Summary and discussion}
\label{sec:discussion}

The main results from this study are the following:

\begin{itemize}
\item A sample of 3269 early B and 930 O stars with matched Gaia EDR3 astrometry has been constructed that contains objects most likely located in the Carina far arm.  These are contained in a 48 sq.deg region defined by $282^{\circ} < \ell < 294^{\circ}$, $-3^{\circ} < b < 1^{\circ}$. The median distance of the sample of 4199 stars, is inferred from the parallax data to be $\sim$5.8 kpc.  
\item There is evident structure in the distribution of longitudinal proper motions, $\mu_{\ell,*}$, that suggests the far arm OB stars are subject to an, as yet, unapportioned combination of kinematic perturbation away from circular motion with spatial corrugation of the far arm. The $\mu_{\ell,*}$ modulation has a saw-tooth character of an amplitude approaching 1 mas yr$^{-1}$, with each rise occupying a longitude range of 2--3 degrees (200--300 pc at median distance).
\item We find a substantial moving group of over 50 OB stars behind the far Carina arm at $\ell \simeq 286^{\circ}$, $b \simeq -1^{\circ}.4$.  Its distance is estimated to be between 7 and 9 kpc: its spatial relation with the far arm, around 6 kpc distant at the same longitude, is unclear at present. 
\item The $\mu_{\ell,\ast}$, $\mu_b$ vector point diagram (Fig.~\ref{fig:pmlvpmb}) shows tight clustering elongated in the longitudinal direction.  This is consistent with the smearing expected from Galactic rotation although it will have suffered some stretch by the observed proper motion modulation.  Using this diagram, a sub-sample of high proper motion outliers containing 288 stars has been identified.  These split into 167 non-emission B stars, 83 non-emission O stars and 38 emission line objects (as classified by MS17 using photometric SED fits).  This translates into 5.7\%, 10.1\% and 10.3\% of the total numbers of non-emission B, O and emission-line stars respectively.  If the selection is based on tangential velocity instead (which needs to use the inferred distance $D_{2.0}$), we obtain very similar percentages if the minimum required velocity is 25 km s$^{-1}$ (244 stars from within a distance-trimmed far-arm sample).   
\item The candidate B stars, \#1$-$04364 and \#1$-$04845, have the most extreme relative proper motions, travelling at tangential velocities in the region of 300 km s$^{-1}$.  Both may be recent ejections from Westerlund 2.
\item The impact parameter method of D19, suitable for finding patterns of runaways around prominent clusters, has been revisited and turned into a wide-area randomisation trial aimed at quantifying the false positive rate.  For searches limited to $\Delta t < 4$~Myr, by-chance candidate counts of up to 3 are possible, depending on how loosely the acceptance criteria are drawn (Table~\ref{tab:stats}). A follow-on simulation without randomisation underlines the pre-eminence of NGC 3603 and Westerlund 2 as sources of runaways in the region. 
\item The results of D18 and D19 for Westerlund 2 and NGC 3603 respectively have been revisited and largely upheld.  New runaway candidates have been identified for both clusters, adding to the pattern of ejections around both of them.  In the case of Westerlund 2, the new additions dilute the strongly bipolar pattern noted by D18.  It is confirmed for both clusters that the detected ejection activity favours times of flight below 1 Myr.  Indeed we find a strong burst of ejections in both only lasting $\sim$300\,000 years. 
\end{itemize}

    
  A pressing question is what has created the marked $\mu_{\ell, * }$ modulation with Galactic longitude seen in Fig.~\ref{fig:pmsvlong}.  It was argued in Section~\ref{sec:structure2} that it is not easily reconciled with pure circular motion around the disk, and must therefore imply the presence of some kinematic perturbation.  A new study of OB stars in Cygnus, $\lesssim 2$ kpc distant, by \cite{Quintana21} has revealed analogous $\mu_{\ell,*}$ modulation that arises unambiguously in perturbed kinematics (see their figure 11).   
  Given how young massive OB stars are, the cause must lie in recent star formation.  
  
  It has been argued by \cite{CKD2020}, from a sample of nearby disc galaxies, that GMCs finally disperse, revealing a new-born stellar population on a typical timescale of $\sim5$ Myr, thanks to the action of HII regions within them promoting feedback outflow velocities averaging around 15 km s$^{-1}$.  As these velocities are half-amplitudes, the scale involved is comparable with the maximum range in velocities implied by the $\mu_{\ell,*}$ modulation found here ($\sim$30 km s$^{-1}$), and also by \cite{Quintana21} who cite a full amplitude for the Cygnus region of 25 km s$^{-1}$.  The angular scale also fits, in that the saw-tooth modulation has a 'wavelength' of 2 degrees or so in Galactic longitude: this converts at the mean distance to the far arm of $\sim$6 kpc to lengths of $\sim$200 pc -- consistent with the scales of 100 to 300 pc identified by \cite{CKD2020}.   Recent modelling reinforces the view that GMC dispersal as observed in the Milky Way responds particularly to the UV radiation flowing from embedded HII regions \citep{Rathjen21}.  In this scenario, the OB-star speeds should not be comparable with the speed of gas outflow -- leaving the option that the OB stars themselves formed in GMC-wide pre-existing organised velocity fields.

The modulation seen -- interpreted purely as kinematic perturbation -- may be analogous to a 'Hubble flow' (expansion in which speed rises with distance from the flow origin).  In the part of the far arm that includes NGC 3603 ($290^{\circ}.0 \lesssim \ell \lesssim 292^{\circ}.5$), perhaps an explosive cluster formation scenario as proposed by \citet{Banerjee2014} might be applicable. Improved constraints on the nature of the velocity field present must come from one or both of greater-precision astrometry (better distances) and spectroscopy (radial velocities).
    
\citet{MA2018} have pointed out that 2D searches for runaway stars, based on proper motions alone, most likely count around half of the runaway stars, compared to the full 3D approach incorporating radial velocity information.  The bright sample studied by these authors (which has no overlap with the sample here) gave the result that 5.7\% of O stars are runaways.  In contrast, \cite{Lamb2016} used radial velocities of SMC field O stars to place a lower limit of 11\% on the runaway fraction.  Our 2D searches based on high proper motion and estimated tangential velocity favour figures close to 10\%. If half are being missed then these results point to around 20\% of O stars in our selections being runaways.  This may be a mild overestimate for the reason that the samples used here do not count all the OB stars in the heart of either NGC 3603 (and misses some in Westerlund 2).  Furthermore, we have not labeled stars in the sample investigated as either 'field' or 'cluster' stars.  If we were to attempt these adjustments, the 20\% O-star runaway fraction (i) would drop to $\sim18$\%, if (applying rough estimates) $\sim$100 O stars were assumed to be hidden and uncounted in cluster cores, (ii) would rise to $\sim28$\% if computed as the field-only fraction, after allocating $\sim$200 to the major clusters.  
    
We find that B stars are half as likely as O stars to be runaway candidates, but note that, in absolute numbers, this translates into twice as many B stars in the runaway population as O stars (regardless of how we measure it).  Among the minority of OB stars that exhibit line emission in their spectra, we have the interesting result that they are at least as likely as O stars to be runaways. Taking the fraction to be 10\%, this can be broken down further (removing the O stars and known Wolf-Rayet stars) to indicate that roughly 8\% of Be stars are escapes. 
This result can be compared with the 3D approach of Boubert \& Evans (2018) that combined space astrometry and spectra in constructing a list of 632 solar-neighbourhood Be stars.  They identified 40 high probability runaways (6.3\%), and used Bayesian inference to estimate an overall incidence in their list of (13.1$+$2.6,$-$2.4)\%.  Again, if our 2D result of 8\% should be doubled to account for the missing dimension, it is compatible. 
    
The last phase of this study of the Carina far arm has focused on the specific identification of candidate runaway OB stars.  Our route to this has been through the trajectory impact parameter method introduced in D19 in application to NGC 3603.   We now know that this method is most effective (in the context of the present quality of the EDR3 proper motions) at finding runaways out to radii on sky of around 1 degree, or for times of flight of up to $\sim$4~Myr.  On this scale, our simulations indicate the detections achieved are relatively free of false positives.  
The method will also pick up moving groups, and it is relatively straightforward to separate them from runaway diaspora just by examining the on-sky pattern of the candidates identified.   

Our updated results on NGC 3603 and Westerlund 2 build on the results given by D19 and D18 for these massive young clusters.  The most striking feature of the lists of candidates obtained is the strong clustering of times of flight to within the last one million years. Indeed, for Westerlund 2, there is a marked rush of 11 out of 13 candidates in the range $0.53 < \Delta t < 0.71$ Myr, a period of only 200\,000 years.  A similar phenomenon is at work in NGC 3603, where there are 9 stars (out of 15) falling in the range $0.55 < \Delta t < 0.87$ Myr.  These are mostly but not only in the 'ring' of runaways noticed by D19. As both clusters are thought to be 1--2 Myr old, this finding implies it does not take long for a newly-formed dense high-mass cluster to evolve to the point of relatively frequent runaway ejection.  These prompt ejections will be dynamical in origin.


This survey of proper motions in the far Carina arm, based on Gaia EDR3 data, creates a complex picture.  The sample of 4199 OB stars has struck down the idealisation that they lie within a smooth spiral arm structure tracing disk rotation, in the mean.  Reality overlays a distinct saw-tooth patterning that most likely arises from the star formation process.  The runaway statistics obtained reinforce some recent results on O and emission line stars, and yet it is clear from our trajectory-tracing simulations that cluster ejections within the region (from NGC 3603 and Westerlund 2) can so far only account for a minority of them.  Extrapolating from our discovery of a grouping of $>50$ OB stars behind the main run of the far arm at $\ell \sim 286^{\circ}$, it is reasonable to anticipate that some of the fast-moving 'field' will be due to presently-unrecognised moving groups.

Compared with the progress to date on analogous populations in the Magellanic Clouds, this study amounts to the first, necessary steps within the Milky Way.  The future promises better astrometry with further Gaia data releases, along with the sixth dimension of radial velocities -- post the commissioning of wide-field massively-multiplexed spectrographs such as ESO's 4MOST.  Our vision of the outstandingly OB-star rich far Carina Arm will become ever more three-dimensional.



     .   


\section*{Acknowledgements}


This paper is based on data products from observations made with ESO Telescopes at the La Silla Paranal Observatory -- specifically, the VST Photometric H$\alpha$ Survey of the Southern Galactic Plane and Bulge (VPHAS+, www.vphas.eu; ESO programme 177.D-3023). 
It is also based on data from the European Space Agency mission Gaia (https://www.cosmos.esa.int/gaia), processed by the Gaia Data Processing and Analysis Consortium (DPAC, https://www.cosmos.esa.int/web/gaia/dpac/consortium). 
Use has also been made of data products from the Two Micron All Sky Survey, which is a joint project of the University of Massachusetts and the Infrared Processing and Analysis Center/California Institute of Technology, funded by the National Aeronautics and Space Administration and the National Science Foundation.  Much of the analysis presented has been carried out via TopCat \citep{Taylor2005}.

MM acknowledges the support by the Spanish Ministry of Science, Innovation and University (MICIU/FEDER, UE) through grant RTI2018-095076-B-C21, and the Institute of Cosmos Sciences University of Barcelona (ICCUB, Unidad de Excelencia “Mar\'ia de Maeztu") through grant CEX2019-000918-M. 
NJW acknowledges receipt of an STFC Ernest Rutherford Fellowship (ref. ST/M005569/1) and a Leverhulme Trust Research Project Grant (RPG-2019-379).

Paul McMillan, the referee of this paper, is thanked for comments that have certainly led to improvements in its content.

\section*{Data Availability}

The main data product associated with this paper is directly available as a fits table supplied as supplementary material.  Its contents are outlined in Appendix~\ref{full_list}.  


\bibliographystyle{mnras}

\appendix

\section{Parameters provided in the candidate OB-star sample listing (supplementary materials) }
\label{full_list}

A fits-formatted OB-candidates list of the 4199 stars selected and discussed in this paper is attached as supplementary material.  The list is laid out with 49 columns of data associated with each object.  The information provided includes data copied across from the MS17 catalogue, selected columns from the Gaia EDR3 database, estimated distances and proper motions expressed in Galactic coordinates.  The table below identifies the parameters/columns provided.


\newpage
\onecolumn
\begin{longtable}[h!]{llcl}
\caption{List of columns available in the catalogue, together with the units and brief description of the column content.
}
\\\label{tab:TableAp}
No & Column & Units & Description \\ \hline
1 & ID & None &  VPHAS-OBM-NNNNN, where NNNNN is ordered by Galactic longitude \\
  &    &      &  and M is 1 or 2. \\ 
2 & RA & degrees &  Right Ascension J2000 (VPHAS$+$)\\ 
3 & DEC & degrees &  Declination J2000 (VPHAS$+$)\\ 
4 & GAL\_LONG & degrees &  IAU 1958 Galactic longitude \\ 
5 & GAL\_LAT & degrees &  IAU 1958 Galactic latitude \\ 
6 & u & mag &  VPHAS+ u band \\ 
7 & u\_err & mag &  VPHAS+ u band photometric uncertainty \\ 
8 & g & mag &  VPHAS+ g band \\ 
9 & g\_err & mag &  VPHAS+ g band photometric uncertainty \\ 
10 & r & mag &  VPHAS+ r band \\ 
11 & r\_err & mag &  VPHAS+ r band photometric uncertainty \\ 
12 & i & None &  VPHAS+ i band \\ 
13 & i\_err & None &  VPHAS+ i band photometric uncertainty \\ 
14 & J & mag &  2MASS J band \\ 
15 & J\_err & mag &  2MASS J band photometric uncertainty \\ 
16 & H & mag &  2MASS H band \\ 
17 & H\_err & mag &  2MASS H band photometric uncertainty \\ 
18 & K & mag &  2MASS K band \\ 
19 & K\_err & mag &  2MASS K band photometric uncertainty \\ 
20 & logTeff & None &  Estimated effective temperature from photometric fits (median of posterior) \\ 
21 & logTeff\_eu\_c & None &  Upper uncertainty on logTeff (84th percentile of posterior) \\ 
22 & logTeff\_el\_c & None &  Upper uncertainty on logTeff (16th percentile of posterior) \\ 
23 & A0 & mag &  Estimated extinction, A0, from photometric fits (median of posterior) \\ 
24 & A0\_eu & mag &  Upper uncertainty on A0 (84th percentile of posterior) \\ 
25 & A0\_el & mag &  Upper uncertainty on A0 (16th percentile of posterior) \\ 
26 & Rv & None &  Photometric estimation for the  extinction law parameter \\ 
27 & Rv\_eu & None &  Upper uncertainty on RV (84th percentile of posterior) \\ 
28 & Rv\_el & None &  Upper uncertainty on RV (16th percentile of posterior) \\ 
29 & chi2 & None &  $\chi^2$ value for photometric fit \\ 
30 & EM\_star & None &  Indicates if star is selected as an emission line object \\ 
31 & Notes & None & (Mainly used for alternate names of objects already in the literature) \\ 
32 & EDR3Name & None &  Gaia EDR3 identifier \\ 
33 & Plx & mas &  Gaia EDR3 parallax \\ 
34 & e\_Plx & mas &  Uncertainty in parallax \\ 
35 & pmRA & mas yr$^{-1}$ &  Proper motion in right ascension\\ 
36 & e\_pmRA & mas yr$^{-1}$ &  Uncertainty in proper motion in right ascension \\ 
37 & pmDE & mas yr$^{-1}$ &  Proper motion in declination\\ 
38 & e\_pmDE & mas yr$^{-1}$ &  Uncertainty proper motion in declination \\ 
39 & pmRApmDEcor & mas yr$^{-1}$ &   Cross-correlation term\\
40 & RUWE & None &  Renormalised Unit Weight Error from Gaia \\ 
41 & Gmag & mag &  Gaia G band \\ 
42 & e\_Gmag & mag &  Gaia G band photometric uncertainty \\ 
43 & D\_2.0 & parsec &  Estimated distance (EDSD inversion: offset 0.03 mas, length scale 2000 pc) \\ 
44 & D\_2.0\_p5 & parsec &  Lower bound on distance (5th percentile of posterior) \\ 
45 & D\_2.0\_p95 & parsec &  Upper bound on distance (95th percentile of posterior) \\ 
46 & pml & mas yr$^{-1}$ &  Longitudinal proper motion \\ 
47 & e\_pml & mas yr$^{-1}$ &  Uncertainty in longitudinal proper motion \\ 
48 & pmb & mas yr$^{-1}$ &  Latitudinal proper motion \\ 
49 & e\_pmb & mas yr$^{-1}$ &  Uncertainty in latitudinal proper motion \\  \hline

\end{longtable}


\end{document}